\documentclass[preprint,12pt]{elsarticle}

\usepackage{amssymb}
\usepackage[export]{adjustbox}

\usepackage[%
    colorlinks=true,
    pdfborder={0 0 0},
    linkcolor=red
]{hyperref}
 \usepackage{amsthm}
 
\usepackage{lineno}
\usepackage{float}
\usepackage{amsmath}
\usepackage{array}
\usepackage{booktabs}
\usepackage{subcaption}
\usepackage{siunitx}
\newcommand{\fluenceUnits}[1]{\SI{#1}{n_{eq}\,\cm^{-2}}}
\newcommand{\lumiUnits}[1]{\SI{#1}{\cm^{-2}\,\second^{-1}}}

\journal{NIM-A}

\begin{document}

\begin{frontmatter}


\title{Impact of Neutron Irradiation on LGADs with a Carbon-Enriched Shallow Multiplication Layer: Degradation of Timing Performance and Gain}


\author[a]{E. Navarrete Ramos \footnote{Corresponding author: efren.navarrete@unican.es}}
\author[a]{J. Duarte-Campderros}
\author[a]{M. Fernández}
\author[a]{G. Gómez}
\author[a]{J. González}
\author[b]{S. Hidalgo}
\author[a]{R. Jaramillo}
\author[a]{P. Martínez Ruiz del Árbol}
\author[d]{M. Moll}
\author[a]{C. Quintana}
\author[c]{A. K. Sikdar}
\author[a]{I. Vila}
\author[b]{J. Villegas}


\affiliation[a]{organization={Instituto de Física de Cantabria, IFCA (CSIC-UC)},
            addressline={Av. los Castros},
            city={Santander},
            postcode={39005},
            country={Spain}}

\affiliation[b]{organization={Instituto de Microelectrónica de Barcelona, IMB-CNM (CSIC)},
             addressline={C/ dels Til·lers Cerdanyola del Vallès},
             city={Barcelona},
             postcode={08193},
             country={Spain}}

\affiliation[c]{organization={Indian Institute of Technology Madras},
             addressline={Tamil Nadu},
             city={Chennai},
             postcode={600036},
             country={India}}

\affiliation[d]{organization={Organisation Europénne pour la Recherche Nucléaire, CERN},
             city={Geneva 23},
             postcode={CH-1211},
             country={Switzerland}}

\begin{abstract}
In this radiation tolerance study, Low Gain Avalanche Detectors (LGADs) with a carbon-enriched broad and shallow multiplication layer were examined in comparison to identical non-carbonated LGADs. Manufactured at IMB-CNM, the sensors underwent neutron irradiation at the TRIGA reactor in Ljubljana, reaching a fluence of $\fluenceUnits{1.5e15}$. The results revealed a smaller deactivation of boron and improved resistance to radiation in carbonated LGADs. The study demonstrated the potential benefits of carbon enrichment in mitigating radiation damage effects, particularly the acceptor removal mechanism, reducing the acceptor removal constant by more than a factor of two. Additionally, time resolution and collected charge degradation due to irradiation were observed, with carbonated samples exhibiting better radiation tolerance. A noise analysis focused on baseline noise and spurious pulses showed the presence of thermal-generated dark counts attributed to a too narrow distance between the gain layer end and the p-stop implant at the periphery of the pad for the characterized LGAD design; however, without significant impact of operation performance.

\end{abstract}


\end{frontmatter}


\section{Introduction}
\label{sec:intro}

The high-luminosity upgrade of the Large Hadron Collider (HL-LHC) is scheduled to begin in early 2029 and will deliver an integrated luminosity of up to \SI{4000}{\per\femto\barn} over a 10-year period~\cite{Aberle:2749422}. The HL-LHC will operate at a stable luminosity of \lumiUnits{5.0e34}, with a possible maximum of \lumiUnits{7.5e{34}}. The main challenge of the HL-LHC will be the superposition of multiple proton-proton collisions per bunch crossing, known as \emph{pileup}, in a small region. The multiple-collision region will extend to about \SI{50}{\milli\meter} RMS along the beam axis, with an average of \SI{1.6}{collisions\per\milli\meter} and up to \SI{200}{pp} interactions per bunch crossing. In these conditions, disentangling the multiple collisions and correctly associating the reconstructed tracks to their primary production vertex will be a major challenge. To address this, MIP timing sub-detectors have been 
proposed~\cite{CERN-LHCC-2017-027, CERN-LHCC-2018-023}, which are targeting a track resolution of \SI{30}{\pico\second} per track. These detectors are expected to significantly improve the performance of the ATLAS and CMS detectors by disentangling the high number of pileup events.

The CMS Endcap Timing Layer (ETL) is a sub-detector proposed to be built using Low Gain Avalanche Detector (LGAD) devices with a pixel size of \qtyproduct{1.3 x 1.3}{\milli\meter\squared}. The ETL will cover the pseudorapidity range of $1.6<|\eta|<3.0$, with a total surface area of \SI{14}{\meter\squared}. This sub-detector will be exposed to radiation levels up to \fluenceUnits{1.5e15} at $|\eta|=3.0$. However, for \SI{80}{\percent} of the ETL area, the fluence is less than \fluenceUnits{1e15}. Therefore, these two fluence points are the ones of interest for this radiation tolerance study.

\begin{figure}
     \centering
         \includegraphics[width=0.73\textwidth]{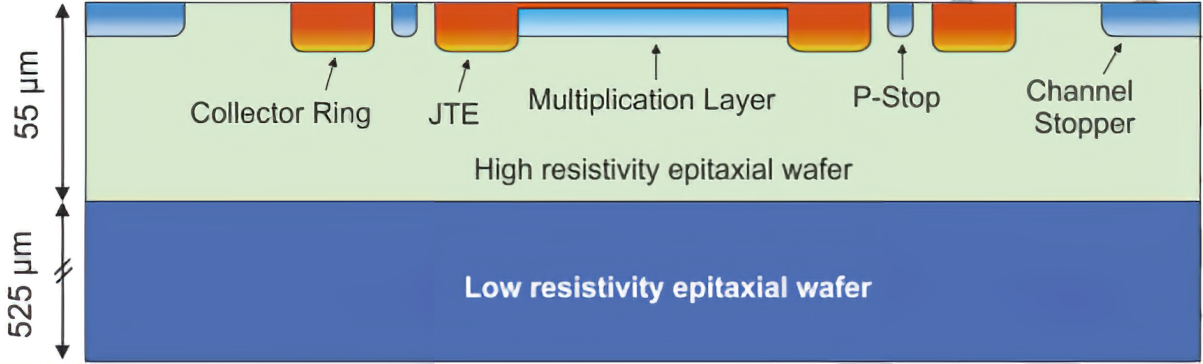}
       \caption{Transversal structure of a Low Gain Avalanche Detector. The multiplication Layer, Collector Ring, P-Stop, Channel Stopper and the Junction Termination Extension (JTE) are visible. Thickness of the active volume and the low resistivity wafer are not scaled.}
        \label{fig:scheme}
\end{figure}

LGADs are semiconductor detectors designed for timing applications. They are constructed as $n^{++}-p^+-p$ avalanche diodes, with a highly-doped $p^{+}$ layer introduced to establish a region of very high electric field. This electric field is responsible for initiating the avalanche multiplication of primary electrons, generating additional electron-hole pairs. A schematic cross-section of a standard pad-like LGAD is illustrated in \autoref{fig:scheme}. The LGAD structure is carefully engineered to achieve a moderate gain and operate effectively across a wide range of reverse bias voltages before reaching breakdown.

We present a radiation tolerance study  performed  on LGAD with a carbon-enriched  multiplication layer. The LGAD sensors were manufactured at IMB-CNM (Institute of Microelectronics of Barcelona, Spain)~\cite{CNM}, with the same processing as the one used in the run \#12916~\cite{Curras}\footnote{The sensor production mentioned in this reference has been developed without carbon enrichment.}. The LGADs are designed with a shallow gain layer doping profile which is characterized by a maximum of the $p^{+}$ doping concentration in the region close to the $n^{++}/p^{+}$ junction and a relative broad  $p^{+}$ implant~\cite{RT-APD}.

Its performance was compared against LGADs with identical layout and manufacturing processing, but without carbon enrichment. The LGADs were irradiated with neutrons at the TRIGA reactor in Ljubljana up to a fluence of \fluenceUnits{1.5e15}.
The degradation of its timing performance and charge collection with fluence is reported.

\section{Samples Description}
\label{sec:descript}

The IMB-CNM carried out a dedicated LGAD production  for studying the effects of carbon on the broad multiplication layer that has been used up to now in its LGAD productions. The LGADs  were manufactured on 6-inch diameter epitaxial wafers with \SI{55}{\micro\meter} active layer thickness and \SI{525}{\micro\meter} support wafer thickness. The handle wafer has a resistivity of \SIrange{0.001}{1}{\ohm\centi\meter} and the substrate resistivity is about \SI{2}{k\ohm\cm}. This run was carried out on epitaxial wafers (Run\#15246, 6LG3 process), and is the first run in which the carbon enrichment of the gain layer was implemented by co-implantation as described in~\cite{Villegas}. The run included matrices of different number of pads, 1$\times$1 (single diodes), 2$\times$2, 5$\times$5, 16$\times$16 and 16$\times$32 where each pad is \qtyproduct{1.3 x 1.3}{\mm\squared}.

Results shown in this work refer to wafers W8 and W10. The manufacturing parameters of these two wafers are described in ~\autoref{tab:wafers} including the gain layer depletion voltage measured before dicing the wafer, the boron dose and the Dry Oxidation Time (DOT). It is important to mention that the main difference between these two wafers is the implementation of carbon to the gain layer of wafer W8 in contrast with W10 that has a standard configuration consisting solely of boron implantation. Employing carbon co-implantation on the gain layer for the LGAD sensors manufacturing is beneficial in reducing the effects of the acceptor removal mechanism~\cite{Ferrero2019} which is an indicator of the degradation of the gain caused by radiation damage. In ~\autoref{sec:acceptor} we show results on the acceptor removal impact on these carbonated sensors.

\begin{table}[htbp]
\centering
\caption{\label{tab:wafers} Differences in gain layer depletion voltage and carbon and boron doses for the carbonated and standard devices}
\smallskip
\begin{tabular}{m{15em}m{2cm} m{2cm} }
\hline
Wafer & Carbonated & Standard\\
\hline\hline
Gain layer depletion Voltage & 30 & 30\\

Boron dose (\SI{1e13}{\per\cm\squared})& 1.9 & 1.9\\

Carbon dose (\SI{1e13}{\per\cm\squared})& 10 & -\\

Dry oxidation time DOT (min) & 180 & 180\\
\hline
\end{tabular}
\end{table}

Measurements were conducted at Instituto de Física de Cantabria (IFCA) in order to characterize sensors from both wafers. ~\autoref{tab:rs-sensors} shows a summary of the sensors measured in the radioactive source setup. Two carbonated and three non-carbonated (standard) were kept as reference. For radioactive source measurements, samples are arranged in stacks of 3 sensors, where one non-irradiated sensor from W10 was used as a reference. Sensors were irradiated with neutrons to 3 different fluences: \fluenceUnits{0.6e15}, \fluenceUnits{1.0e15} and \fluenceUnits{1.5e15}, in the \SI{250}{\kilo\watt} TRIGA Mark II reactor\footnote{Capable of yield a maximum flux of around \SI{2e13}{n\,\cm^{-2}\,\second^{-1}}~\cite{TRIGA}.} of the Jožef Stefan Institute (JSI)~\cite{Zontar}  at Ljubljana (Slovenia). The standard sensors at the highest fluence were not available for this study.

\begin{table}[htbp]
\centering
\caption{\label{tab:rs-sensors} Summary of the LGADs measured in radioactive source setup }
\smallskip
\begin{tabular}{m{9em} m{6em} m{6em} }
\hline
Fluence (\fluenceUnits{ }) & Carbonated & Standard \\
\hline\hline
0                       & D207, D253 & D302, D297 \\
$0.6 \times 10^{15}$    & D238, D217 & D228, D259 \\
$1.0 \times 10^{15}$    & D259, D255 & D277, D275 \\
$1.5 \times 10^{15}$    & D315, D269 & -          \\
\hline
\end{tabular}
\end{table}

The total number of sensors measured in electrical characterization is higher than the ones measured in the radioactive source setup as we will see in \autoref{sec:Electric}.

\section{Electrical Characterization}
\label{sec:Electric}

The Current-Voltage (IV) and Capacitance-Voltage (CV) characteristics were measured in a probe station equipped with a thermal chuck. The measurements were performed before and after irradiation. Measurements of non-irradiated devices were conducted at room temperature, while the irradiated devices were measured at a temperature of \SI{-25}{\celsius}. The ohmic contact side (backside) of the sensors was connected to ground, while the cathode and the guard-ring were connected to High-Voltage (HV). For the IV measurement, the guard-ring and main diode currents were determined independently using two different Keithley 2410 sourcemeters~\cite{keith} that allows supply the High-Voltage for the diode and measure the currrent at the same time. 
For the CV measurement, the guard-ring and the main diode were connected to HV using Keithley 2410 sourcemeters and read by a Quadtech 1920 LCR-meter~\cite{LCR} connected through a decoupling box. The circuit model used to determine the capacitance was a parallel RC circuit and the measurements were carried out at \SI{1}{\kilo\hertz} (\SI{100}{\hertz}) frequency before (after) irradiation. In total, 44 LGADs were measured in IV and approximately half of them in CV before irradiation.

\subsection{Current-Voltage characteristic}
\label{sec:iv}

\begin{figure}
     \centering
     \begin{subfigure}[b]{0.49\textwidth}
         \centering
         \includegraphics[width=\textwidth]{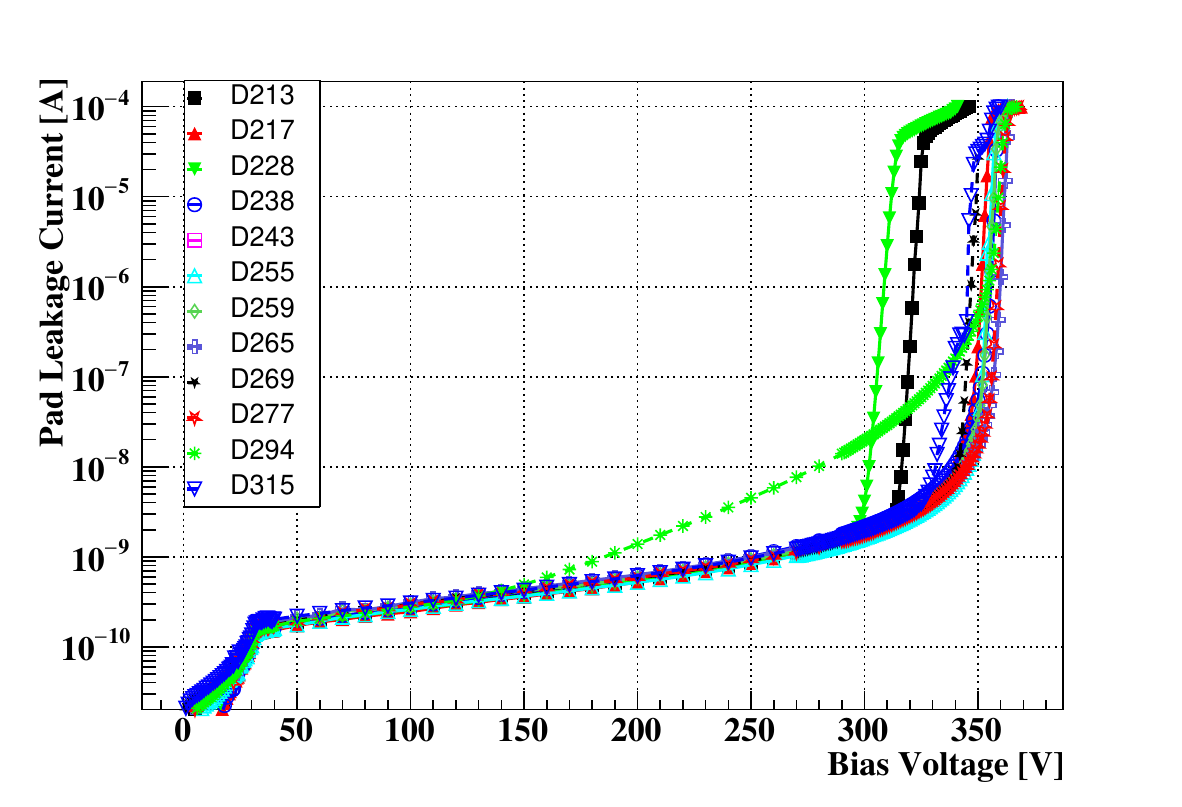}
         \caption{Carbonated}
     \end{subfigure}
     \hfill
     \begin{subfigure}[b]{0.49\textwidth}
         \centering
         \includegraphics[width=\textwidth]{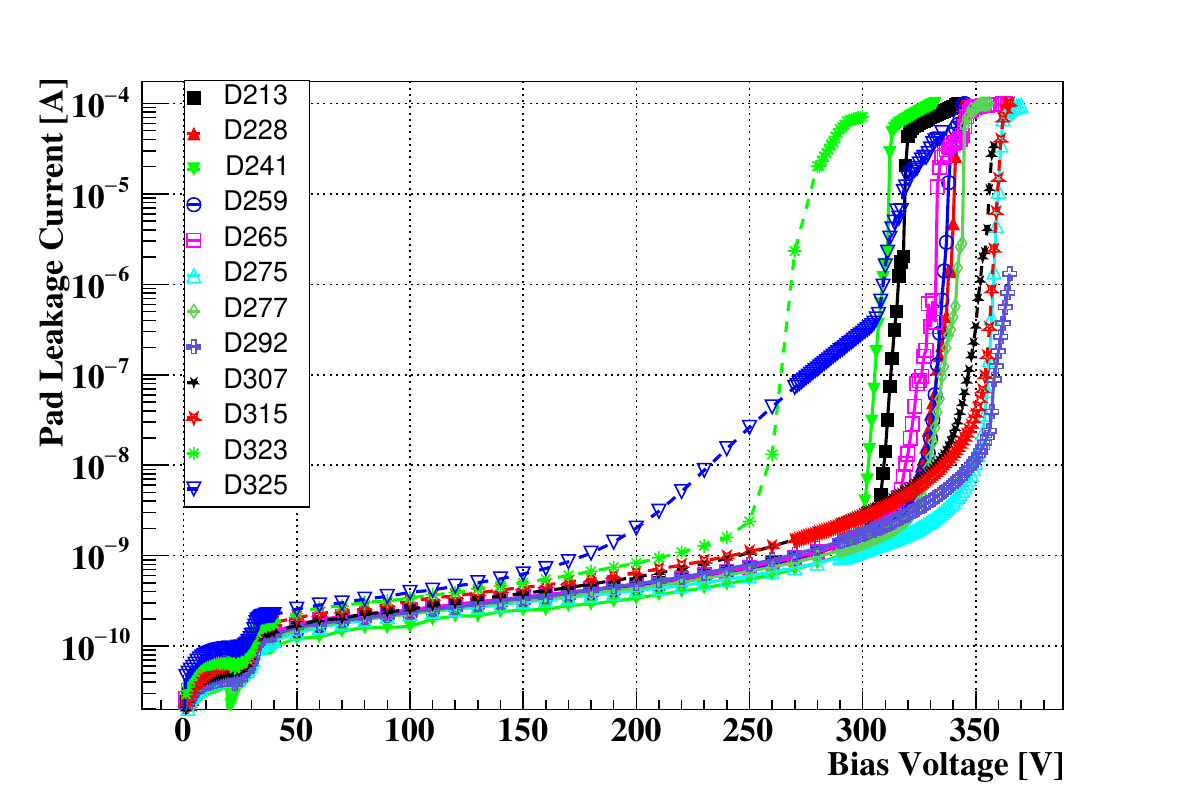}
         \caption{Standard}
     \end{subfigure}
     \centering
     \begin{subfigure}[b]{0.49\textwidth}
         \centering
         \includegraphics[width=\textwidth]{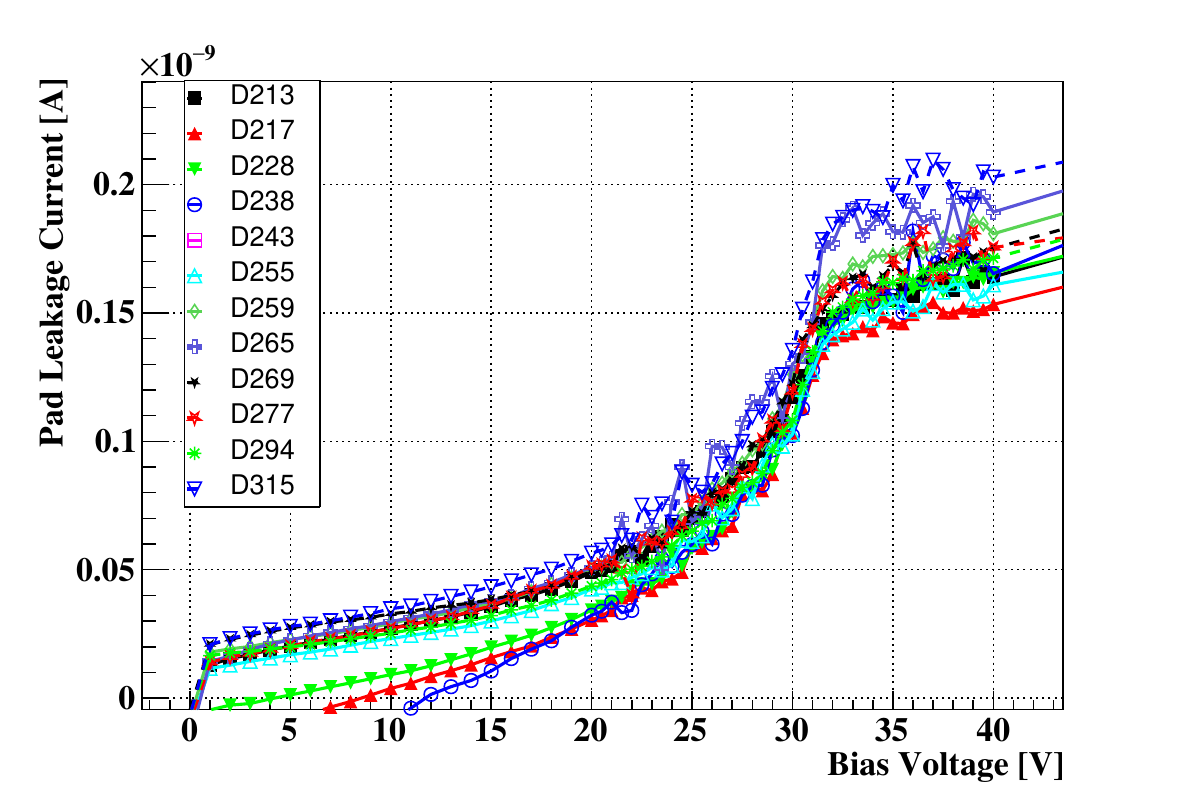}
         \caption{Carbonated GL region}
     \end{subfigure}
     \hfill
     \begin{subfigure}[b]{0.49\textwidth}
         \centering
         \includegraphics[width=\textwidth]{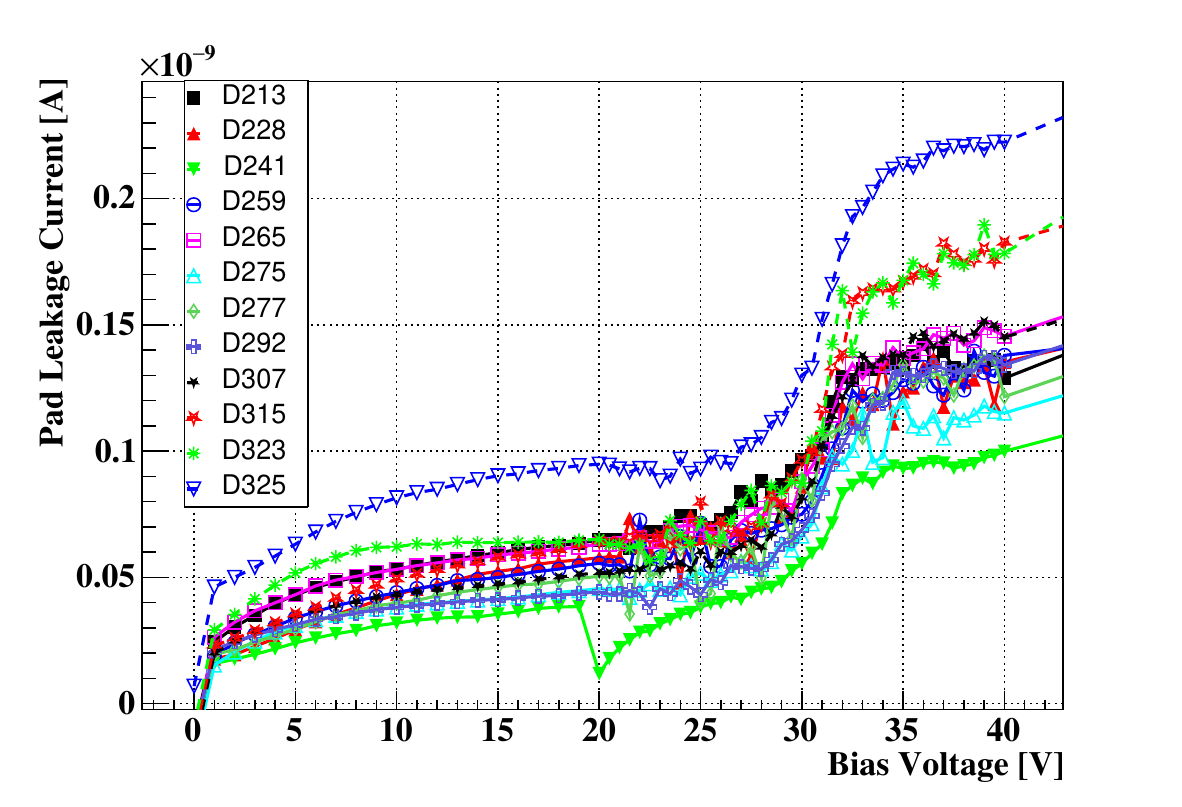}
         \caption{Standard GL region}
     \end{subfigure}
        \caption{The leakage currents of the main diode are presented as a function of reverse bias before irradiation. (a) shows the results for carbonated sensors, (b) shows the results for standard sensors. (c) and (d) exhibit a zoomed-in view of the bias region where the gain layer is depleted.}
        \label{fig:IV_pre}
\end{figure}

\autoref{fig:IV_pre} displays the main diode current versus the reverse bias at room temperature for both types of sensors before irradiation. Across most of the bias voltage range, the current for both types of sensors is below the nanoampere. At full depletion, for example at \SI{250}{\volt}, the leakage current for the carbonated sensors is around \SI{0.91}{\nano\ampere}, while for the standard sensors it is about \SI{0.89}{\nano\ampere}, representing a small difference of approximately $2\%$.

The breakdown voltage $V_{BD}$ was determined for both carbonated and standard sensors estimating the change in the slope by using the method described in~\autoref{sec:acceptor}. A modest increase in the breakdown voltage is observed for the carbonated samples compared to the standard ones. The $V_{BD}$ was found to be in the range of \SIrange{290}{350}{\volt} for the carbonated samples and \SIrange{280}{340}{\volt} for the standard samples. The average $V_{BD}$ is \SI{332}{\volt} for the carbonated sensors and \SI{321}{\volt} for the standard sensors, with a value dispersion (RMS) of approximately \SI{17}{\volt} and \SI{24}{\volt}, respectively.
It is possible that, due to the relatively small size of the carbonated region compared to the remaining non-carbonated silicon active bulk, not much impact of the carbon on the $V_{BD}$ may be expected.

\begin{figure}
     \centering
     \begin{subfigure}[b]{0.49\textwidth}
         \centering
         \includegraphics[width=\textwidth]{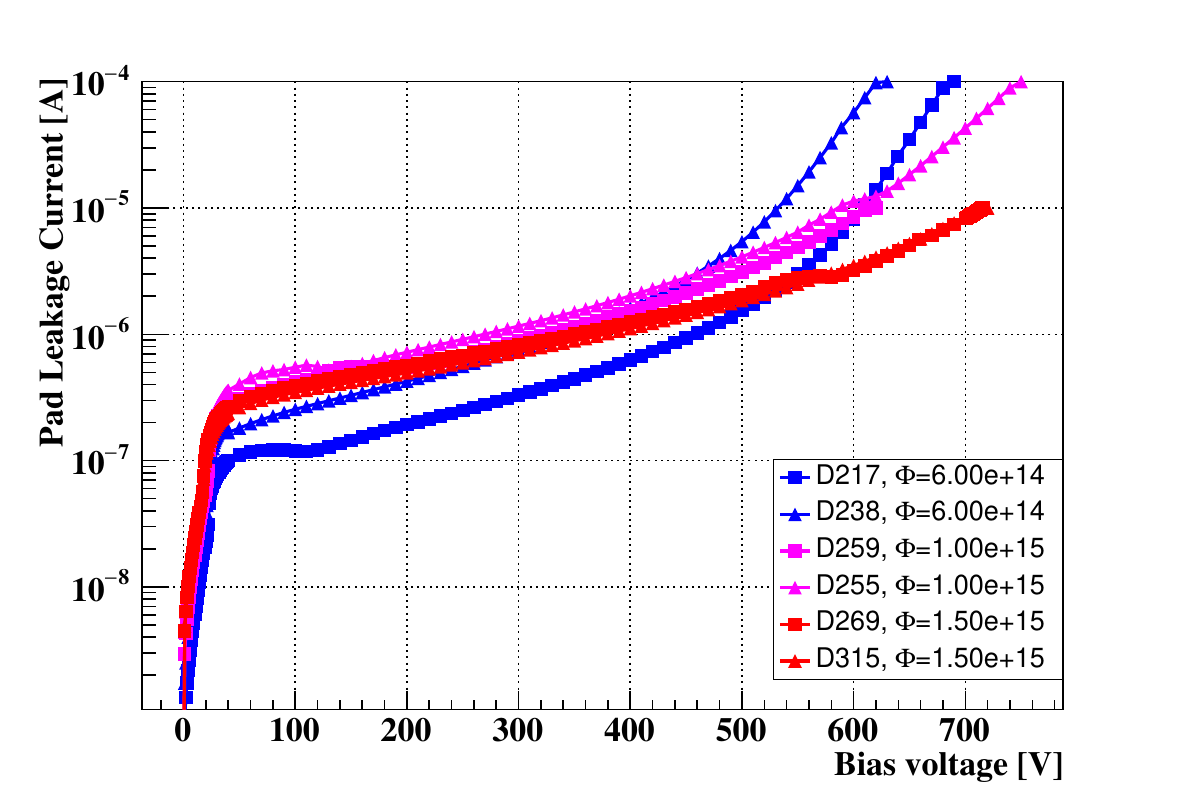}
         \caption{Carbonated irradiated}
     \end{subfigure}
     \hfill
     \begin{subfigure}[b]{0.49\textwidth}
         \centering
         \includegraphics[width=\textwidth]{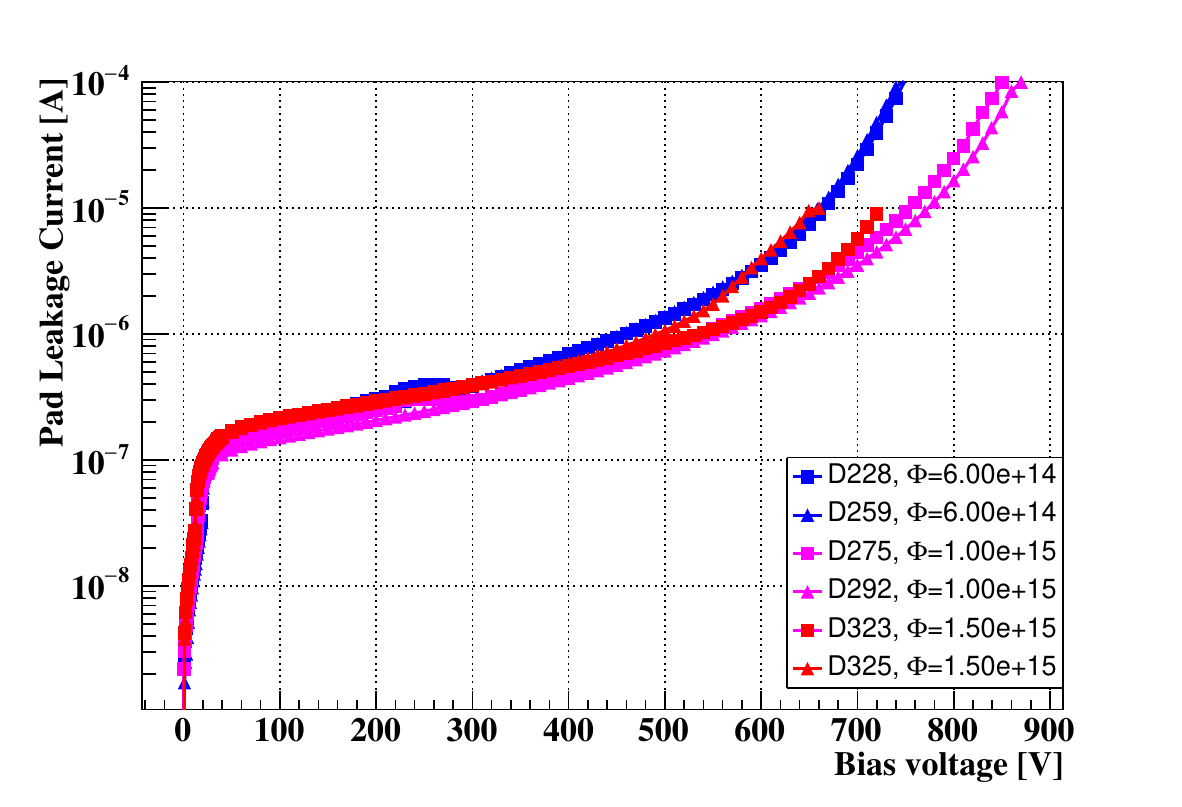}
         \caption{Standard irradiated}
     \end{subfigure}
        \caption{Pad leakage currents after irradiation as a function of the reverse bias. The carbonated sensors are presented in (a) and the standard in (b). These IV curves are shown in log scale for Y axis.}
        \label{fig:IV_irr}
\end{figure}

A second electrical characterization at \SI{-25}{\celsius} was carried out after irradiation of the devices, from which two sensors of every type and fluence (\fluenceUnits{0.6e15} and \fluenceUnits{1.5e15}) were characterised. In \autoref{fig:IV_irr} (a) the pad leakage current of the carbonated sensors can be observed as a function of the reverse bias. We did not observe a sharp breakdown in the IV characteristics for the irradiated samples. Instead, the IV curves show a gradual rise in current, indicating a soft breakdown as the bias increases. We avoided applying a bias high enough to reach full breakdown because we did not want to operate the sensor beyond the single event burnout~\cite{SEB} limit.

For both carbonated and standard samples, we did not observe a monotonic increase in the leakage current with fluence, suggesting the presence of radiation-induced gain suppression. This gain reduction appears to compensate for the radiation-induced increase in leakage current.


\subsection{Capacitance-Voltage characteristic}
\label{sec:cv}

\begin{figure}
     \centering
     \begin{subfigure}[b]{0.49\textwidth}
         \centering
         \includegraphics[width=\textwidth]{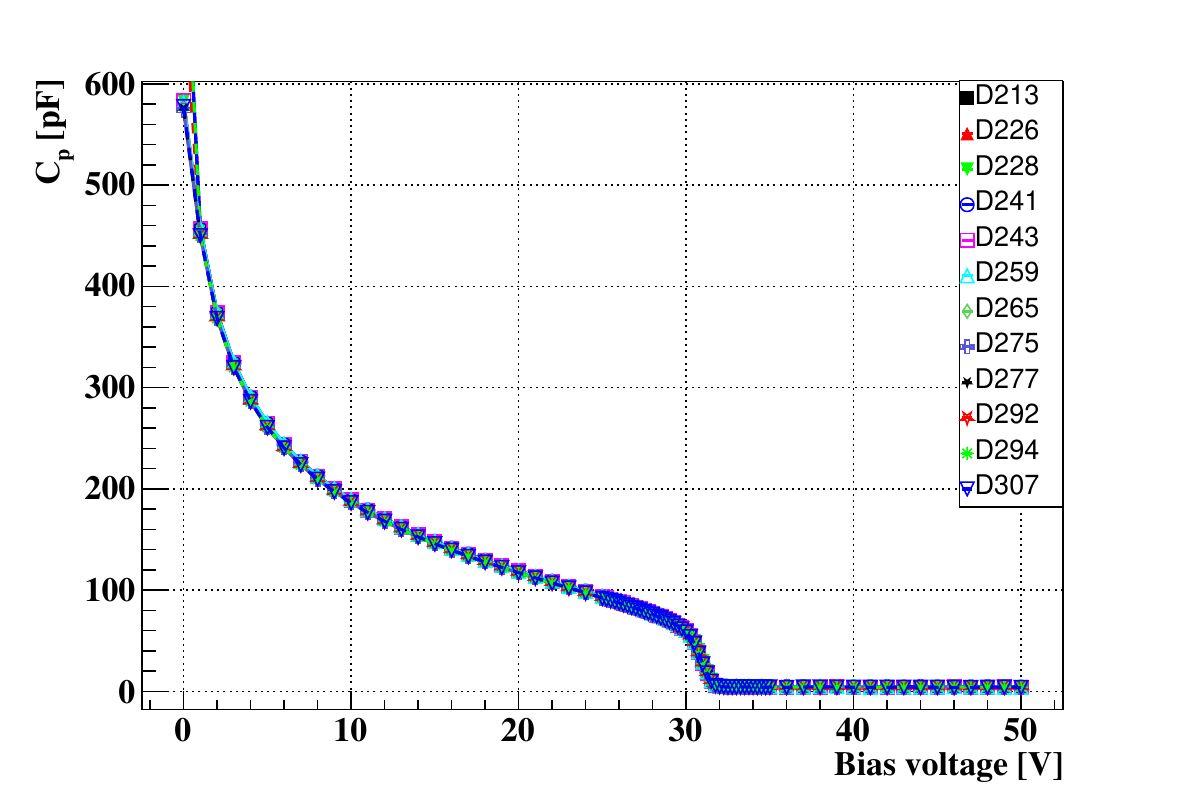}
         \caption{Carbonated}
     \end{subfigure}
     \hfill
     \begin{subfigure}[b]{0.49\textwidth}
         \centering
         \includegraphics[width=\textwidth]{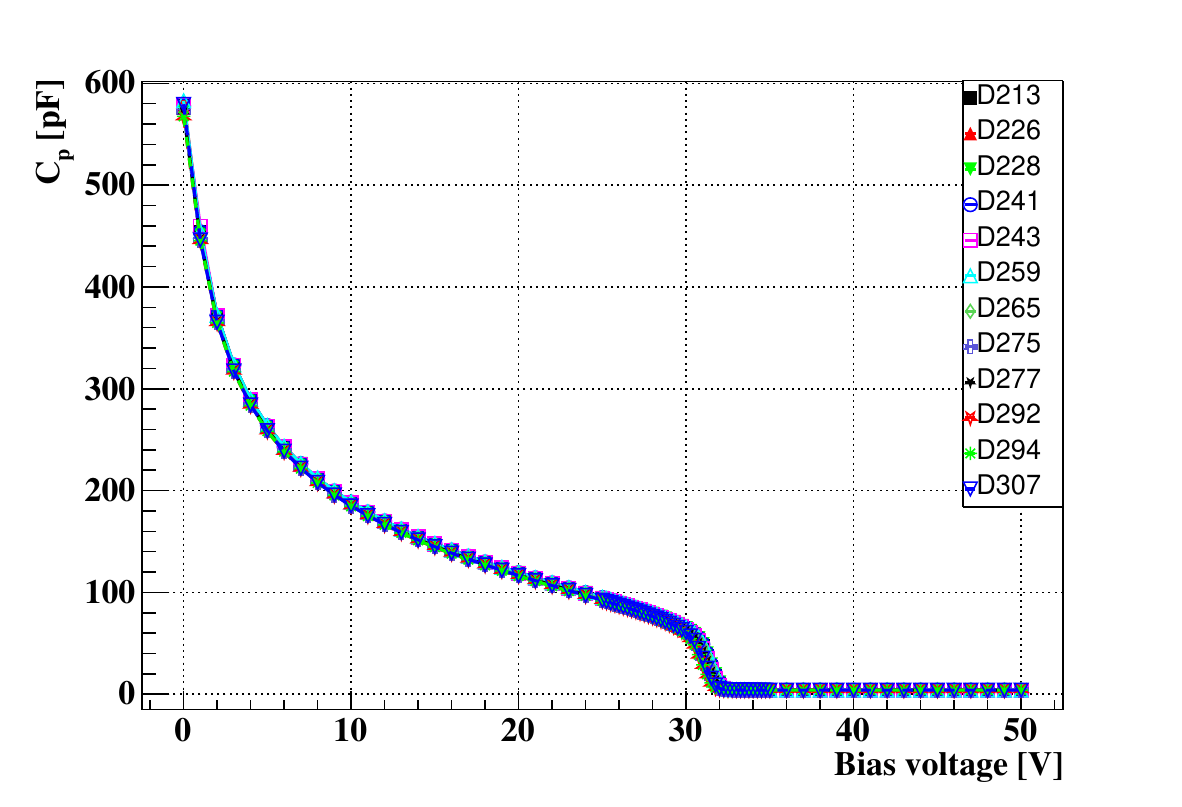}
         \caption{Standard}
     \end{subfigure}
     \begin{subfigure}[b]{0.49\textwidth}
         \centering
         \includegraphics[width=\textwidth]{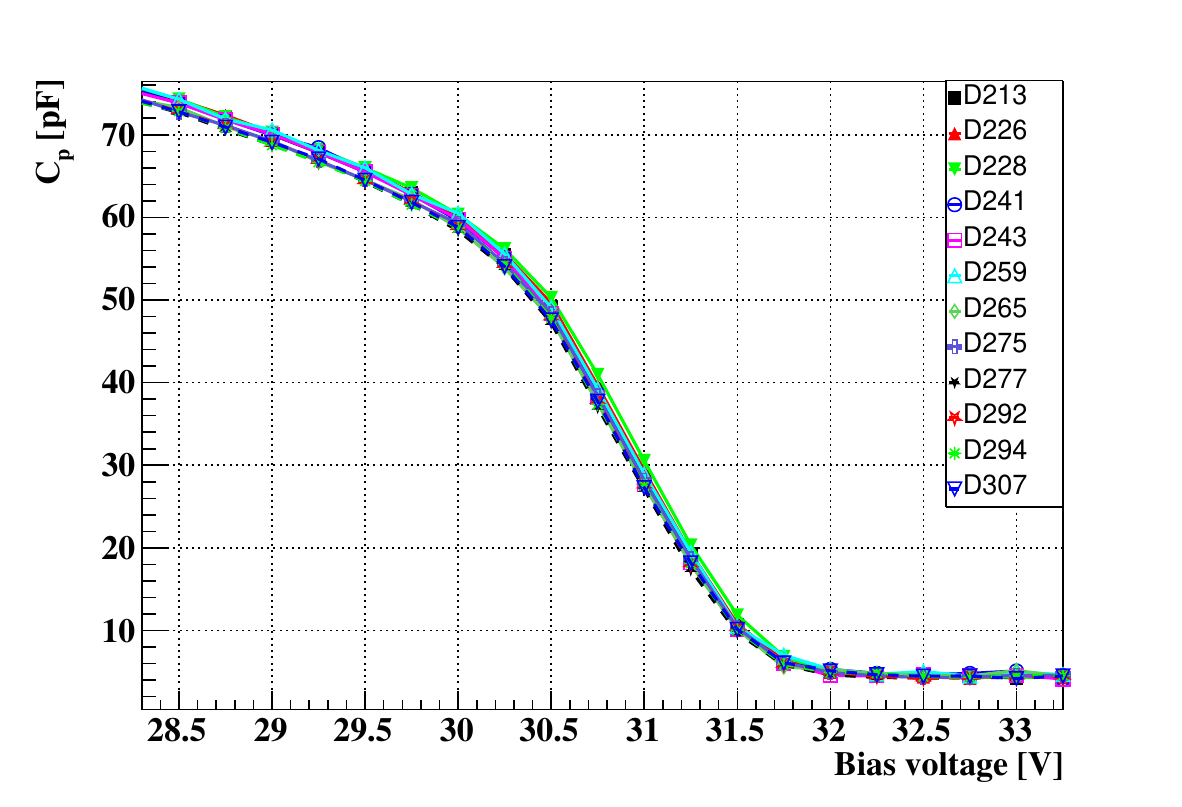}
         \caption{Carbonated GL region}
     \end{subfigure}
     \hfill
     \begin{subfigure}[b]{0.49\textwidth}
         \centering
         \includegraphics[width=\textwidth]{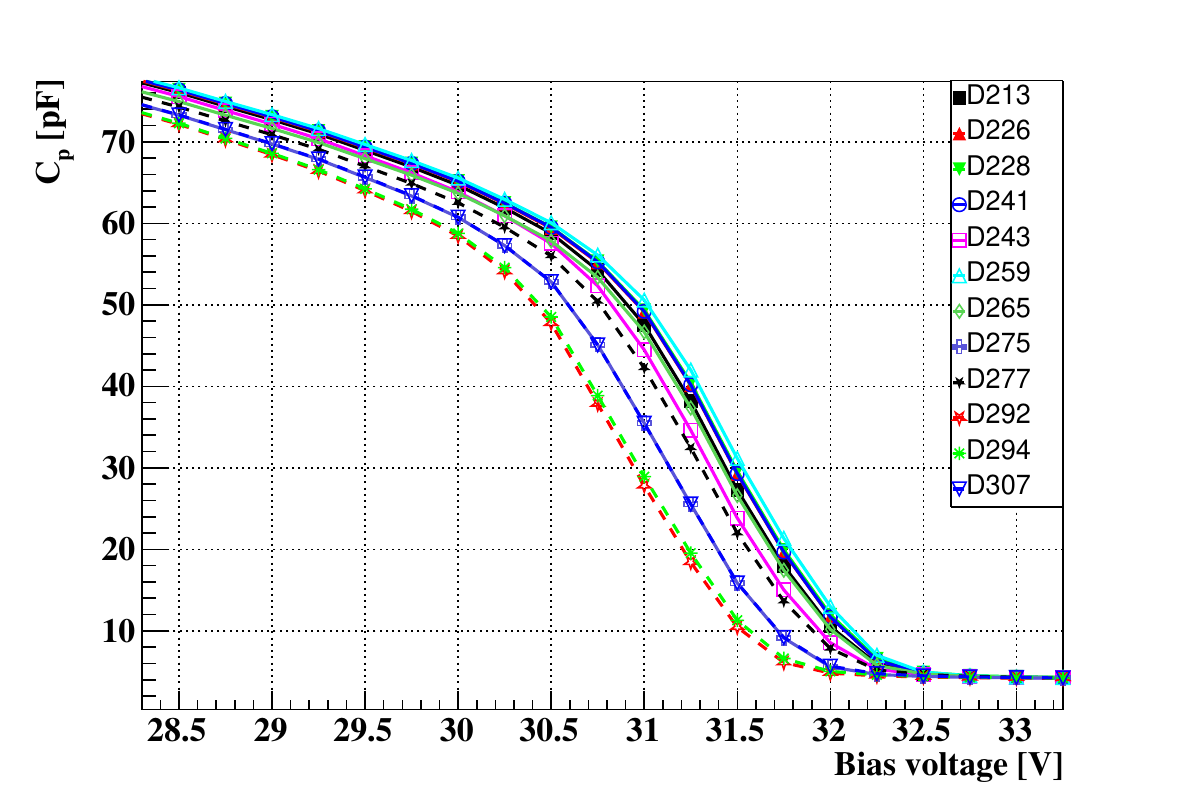}
         \caption{Standard GL region}
     \end{subfigure}
        \caption{Pad capacitance before irradiation as a function of the reverse bias. The carbonated sensors from the carbonated wafer are presented in (a) and the standard sensors in (b). (c) and (d) are zoomed views of the gain layer regions. The characteristic kinks in the curves due to the gain layer and bulk depletion can be observed.}
        \label{fig:cv_pre}
\end{figure}

The capacitance of the bare sensors was measured before irradiation at room temperature with the guard-ring connected and at a frequency of \SI{1}{\kilo\hertz} on the LCR-meter. The curves of the capacitance versus the reverse bias applied for the carbonated and standard samples, are shown in \autoref{fig:cv_pre} (a) and \autoref{fig:cv_pre} (b) respectively. High homogeneity and reproducibility are evident in these curves. The CV curves start with a smooth decrease in capacitance in the gain layer region that ends at approximately \SI{30}{\volt} for both carbonated and standard samples, representing the depletion voltage \(V_{GL}\). This is followed by another kink in the curve, indicating the depletion of the bulk, which ends with a final capacitance (\(C_{\text{end}}\)) of about \SI{4.0}{\pico\farad} for both types of wafers, at voltages above \SI{32}{\volt} (carbonated) and \SI{32.5}{\volt} (standard). This \(C_{\text{end}}\) is consistent with the fact that all sensors have the same active area and width. \autoref{fig:cv_pre} (c) and \autoref{fig:cv_pre} (d) show an enlarged view of the capacitance curves in the gain layer region, where it can be observed that, in general, the curves of all samples, carbonated and standard, follow a similar shape, but the \(V_{GL}\) of the samples is less dispersed in the presence of carbon.

To better observe the $V_{GL}$ region~\cite{Campbell}, these measurements were conducted at a temperature of \SI{10}{\celsius} with a low frequency of \SI{100}{\hertz} configured in the LCR-meter. The remaining LCR-meter settings remained consistent with those used before irradiation.

\begin{figure}
     \centering
     \begin{subfigure}[b]{0.49\textwidth}
         \centering
         \includegraphics[width=\textwidth]{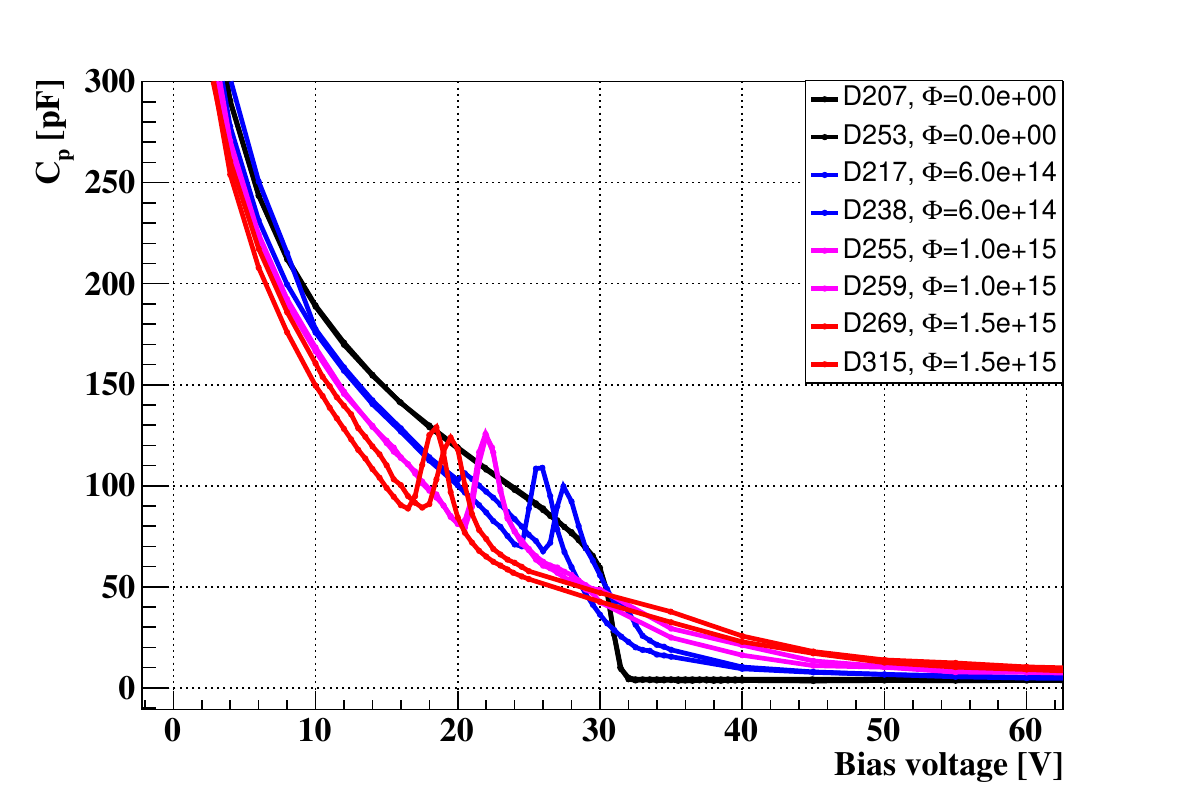}
         \caption{Carbonated}
     \end{subfigure}
     \hfill
     \begin{subfigure}[b]{0.49\textwidth}
         \centering
         \includegraphics[width=\textwidth]{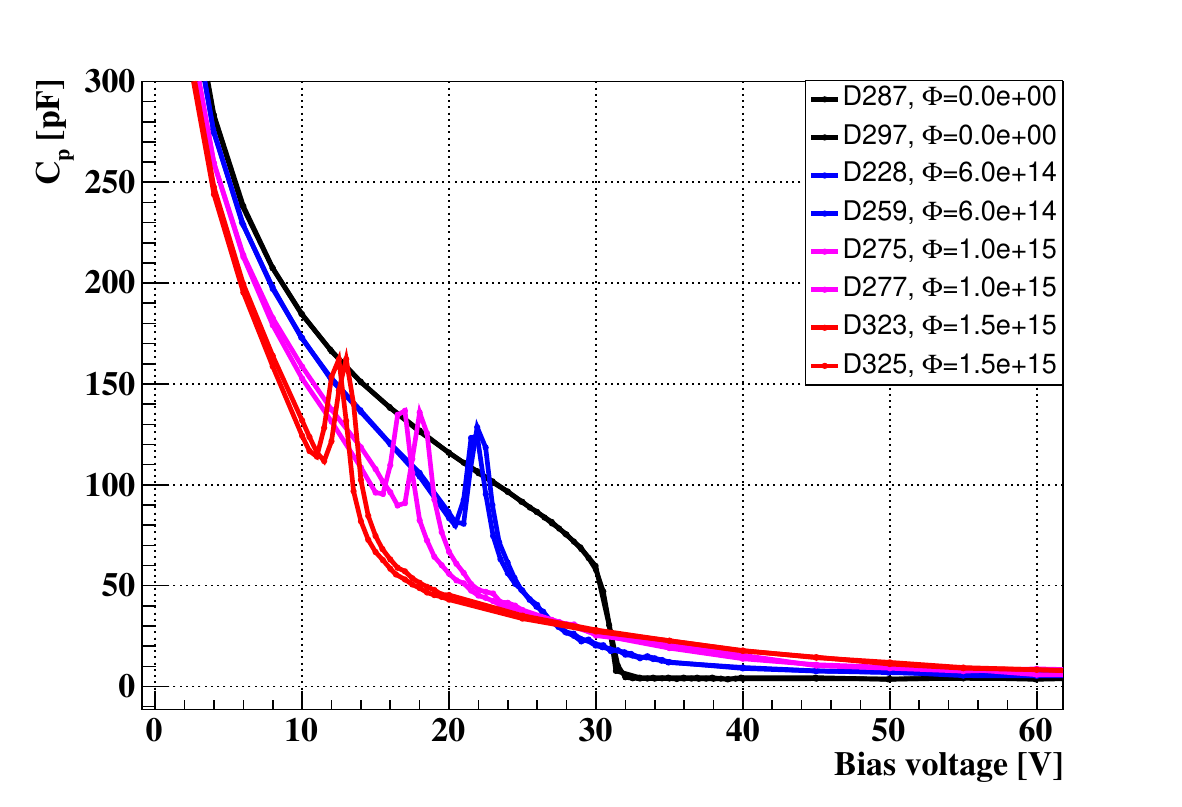}
         \caption{Standard}
     \end{subfigure}
        \caption{Pad capacitance after irradiation as a function of the reverse bias. Non-irradiated samples added for comparison. The carbonated sensors are presented in (a) and the standard sensors in (b). Displacement of the $V_{GL}$ (start of the peak in the curve) as result of the irradiation at the three different fluences is observed.}
        \label{fig:cv_after}
\end{figure}

The pad capacitance after irradiation of the samples is depicted as a function of the applied bias in \autoref{fig:cv_after} (a) for the carbonated samples and \autoref{fig:cv_after} (b) for the standard devices. After irradiation we can see a peak in the capacitance until a local maximum that has been concluded to be related with the presence of the multiplication layer~\cite{Moritz}. The carbonated samples exhibit a noticeable degradation in the gain layer due to irradiation, resulting in a corresponding shift of $V_{GL}$ proportional to the fluence (lower $V_{GL}$ at higher fluences). The standard samples, also present a reduction in $V_{GL}$ that again is evident as a consequence of irradiation. However, the $V_{GL}$ values of the standard samples are lower compared to the carbonated samples. For instance, at the highest fluence, the $V_{GL}$ for the carbonated sensor is approximately \SI{17}{\volt} while for the standard sensor, it is around \SI{11}{\volt}. From these CV characteristics, we have considered the $V_{GL}$ as the last point before the increase of the capacitance, since this coincides with the $V_{GL}$ extracted from IV characteristics.

\subsection{Determination of Acceptor Removal Coefficient}
\label{sec:acceptor}

It has been shown that LGAD sensors experience a reduction in gain after irradiation with charged hadrons or neutrons~\cite{Galloway2019}. This reduction can be attributed to the initial acceptor removal mechanism, involving the gradual deactivation of acceptors forming the GL~\cite{Kramberger}, specifically boron (B) in this case.

As irradiation deactivates the boron implanted in the GL of LGAD sensors, the reverse bias required to fully deplete this gain layer, denoted as $V_{GL}$, decreases compared to the pre-irradiation state. This reduction in $V_{GL}$ provides an indication of the remaining active boron in the GL. Assuming uniform boron removal throughout the multiplication layer and at a consistent rate, we can express $V_{GL}$ as proportional to the boron concentration using the following equation:

\begin{equation}
\label{eq:acceptor}
V_{GL}(\Phi)\approx V_{GL}(\Phi=0)\times e^{-c\Phi}
\end{equation}

Here, $c$ is the acceptor removal coefficient, and $V_{GL}$ represents the gain layer depletion voltage corresponding to the given fluence $\Phi$. The coefficient $c$ is an indicator of the degradation suffered by the multiplication layer and thus the lower $c$ value, the more radiation hard the sensor is.

\begin{figure}
     \centering
     \begin{subfigure}[b]{0.49\textwidth}
         \centering
         \includegraphics[width=\textwidth]{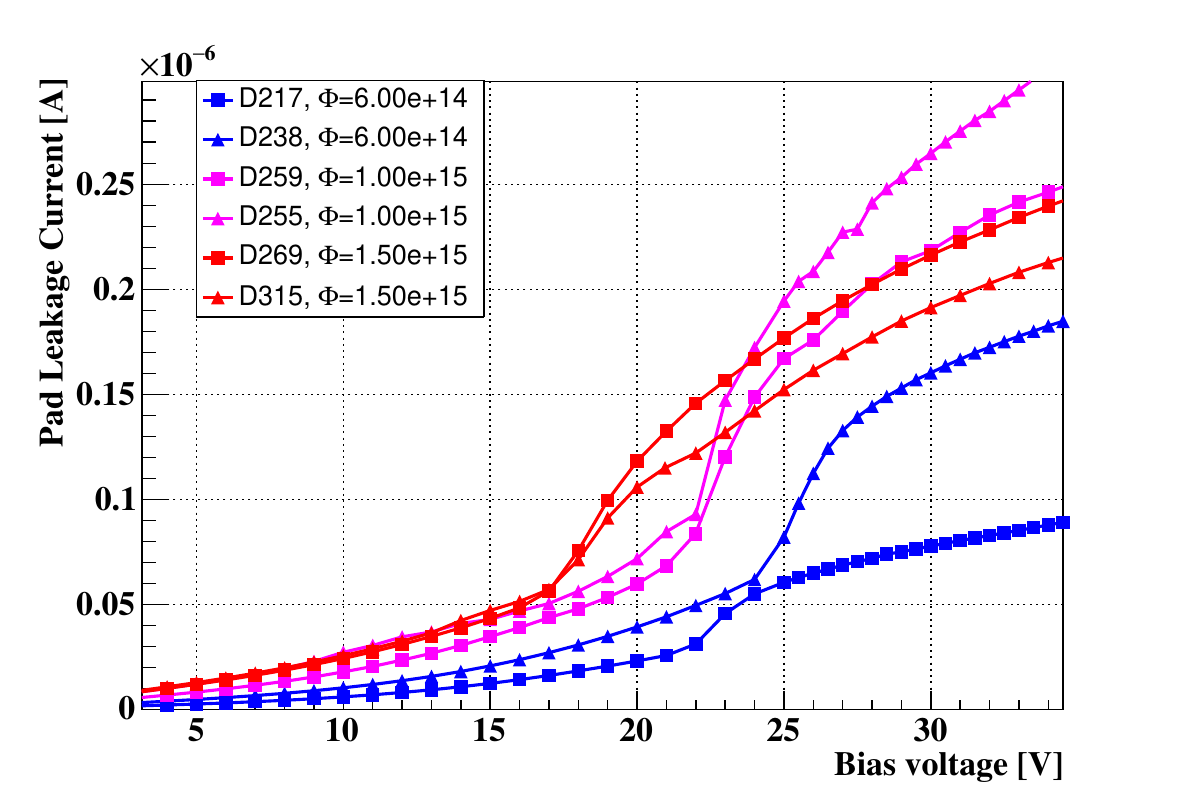}
         \caption{Carbonated}
     \end{subfigure}
     \hfill
     \begin{subfigure}[b]{0.49\textwidth}
         \centering
         \includegraphics[width=\textwidth]{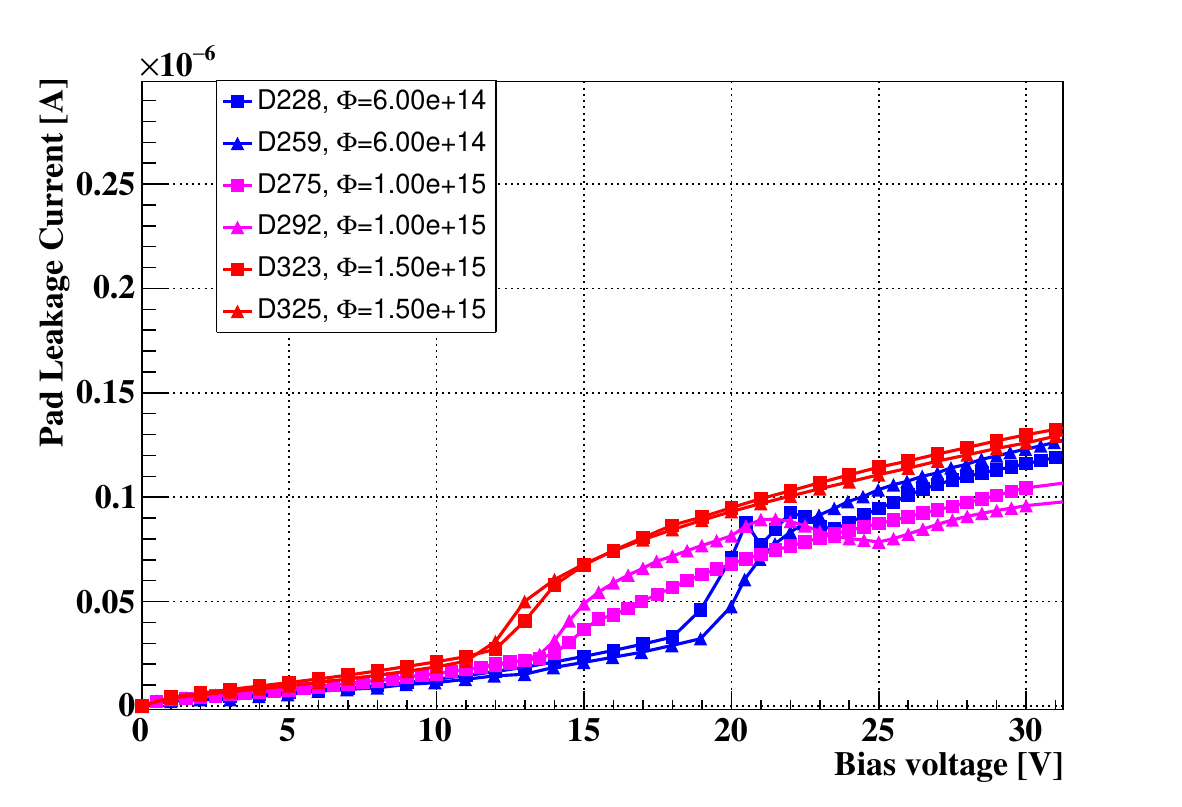}
         \caption{Standard}
     \end{subfigure}
     \begin{subfigure}[b]{0.49\textwidth}
         \centering
         \includegraphics[width=\textwidth]{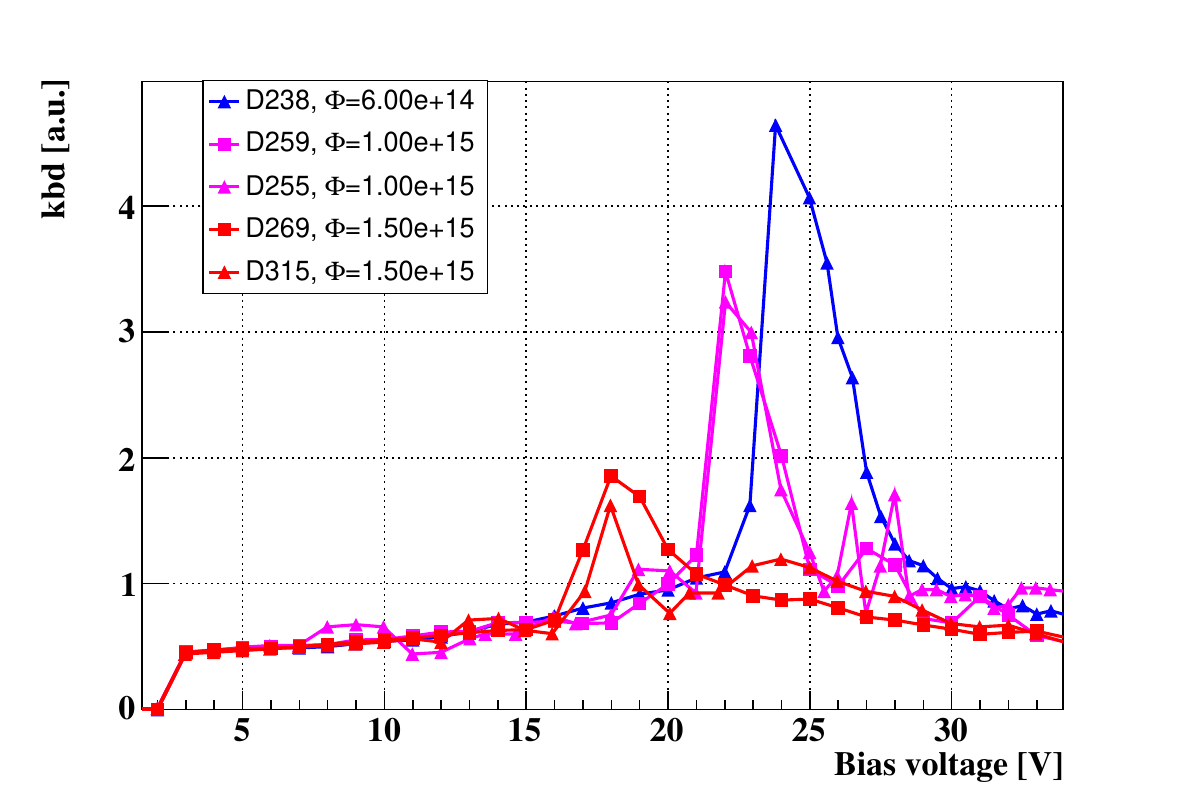}
         \caption{kbd variable, carbonated}
     \end{subfigure}
     \hfill
     \begin{subfigure}[b]{0.49\textwidth}
         \centering
         \includegraphics[width=\textwidth]{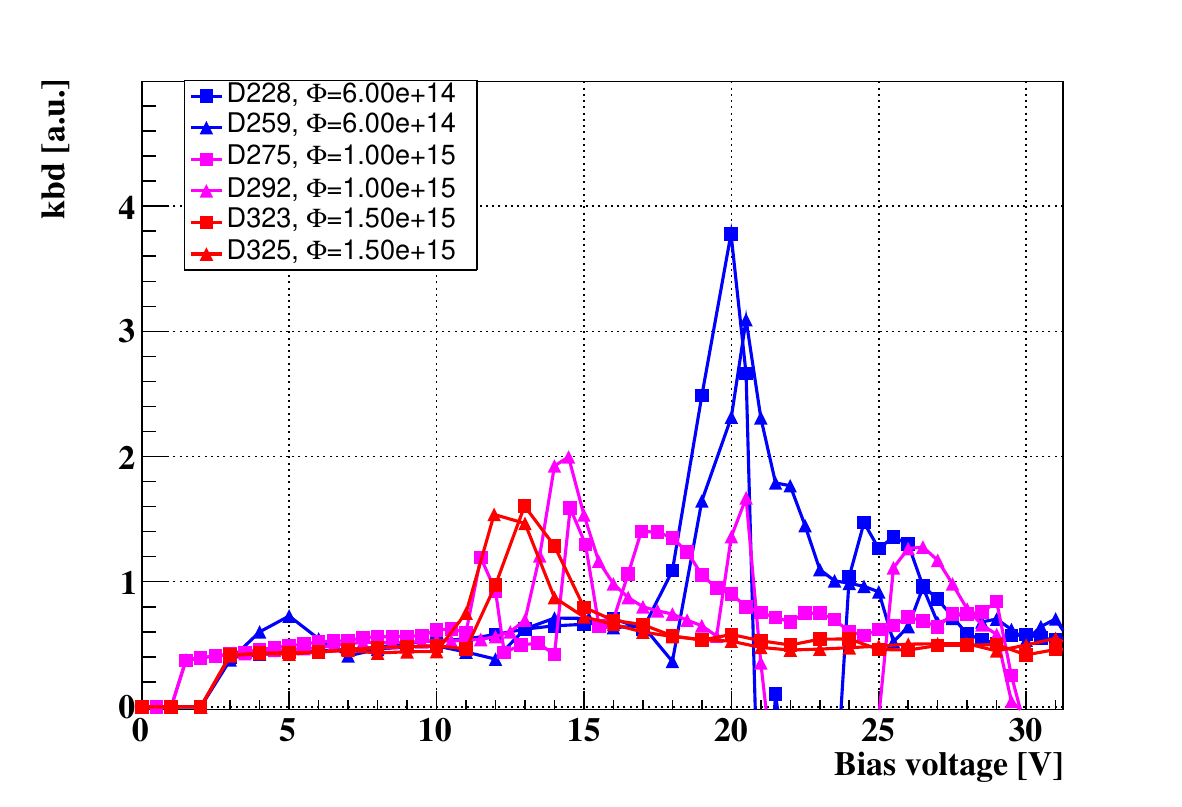}
         \caption{kbd variable, standard}
     \end{subfigure}
        \caption{(a) and (b) presents an enlarged view of the GL regions from the IV curves of sensors at different fluences. (c) and (d) contains the variable kbd stated in \autoref{eq:kbd} and derived from the IV curves, plotted as a function of voltage in order to identify the transition from the GL to the bulk of the sensors.}
        \label{fig:KBD}
\end{figure}

The co-implantation of carbon into the gain layer of LGADs has been proposed to mitigate the adverse effects of radiation~\cite{Kramberger-carbon}, by acting like a trap that prevents the formation of $B_iO_i$ defects, the primary cause of the acceptor removal mechanism~\cite{bi-stable}. These results are also consistent with the evidence of better performance of carbonated sensors against the standard sensors, seen in the improvement in acceptor removal from other LGAD manufacturers~\cite{acceptor-producers}.

After irradiation, a second electrical characterization was conducted to analyze the degradation of the gain layer, starting with the extraction of $V_{GL}$ and determining the Acceptor Removal Coefficient. A crucial aspect of this study is examining the effect of carbon enrichment in the GL compared to the standard boron implantation and how it influences the acceptor removal coefficient for both types of sensors.

\autoref{fig:KBD} (a) and \autoref{fig:KBD} (b) contain the IV curves, centered around the GL region, of the carbonated and standard sensors respectively. \autoref{fig:KBD} (c) and \autoref{fig:KBD} (d) we show a variable constructed as the derivative of the current weighted by the ratio of current over voltage:

\begin{equation}
\label{eq:kbd}
\mathrm{kbd}=\frac{dI/dV}{I/V}
\end{equation}

\begin{table}[htbp]
\centering
\caption{\label{tab:vgls} Summary of the $V_{GL}$ values for both type of sensors, extracted from the electrical characterization (IV and CV) before and after irradiation. The errors are the standard error of the mean (SEM).}
\smallskip
\begin{tabular}{m{8.5em} m{5em} m{5em} m{5em} m{5em}}
\hline
& \multicolumn{2}{c}{$V_{GL}$ from IV (V)} & \multicolumn{2}{c}{$V_{GL}$ from CV (V)}\\
Fluence ($\fluenceUnits{}$) & Carbonated & Standard & Carbonated & Standard \\
\hline\hline
0 & $30.7\pm0.2$ & $29.1\pm0.2$ & $30.5\pm0.1$ & $30.5\pm0.2$ \\
$0.6\times 10^{15}$ & $23.9\pm0.9$ & $19.4\pm0.2$  & $25.3\pm0.2$  &  $20.8\pm0.5$ \\
$1.0\times 10^{15}$ & $22.3\pm0.5$ & $14.2\pm0.7$  & $20.5\pm0.3$  &  $15.7\pm0.2$  \\
$1.5\times 10^{15}$ & $17.5\pm0.5$ & $12.8\pm0.5$  & $17.2\pm0.2$  & $11.3\pm0.2$  \\
\hline
\end{tabular}
\end{table}

This variable was first introduced in~\cite{Bacchetta} as an automatic estimator of the breakdown voltage and is used here to identify the change in slope indicating the transition from the GL to the bulk~\cite{Marcos}\footnote{Other methods to calculate the $V_{GL}$, see: \cite{Vagelis}}. The $V_{GL}$ values extracted from the electrical characterization are shown in~\autoref{tab:vgls}. From these values the degradation of the GL can be calculated fitting the dependence of $V_{GL}$ with fluence, according to~\autoref{eq:acceptor}.

\begin{figure}
     \centering
     \begin{subfigure}[b]{0.43\textwidth}
         \centering
         \includegraphics[width=\textwidth]{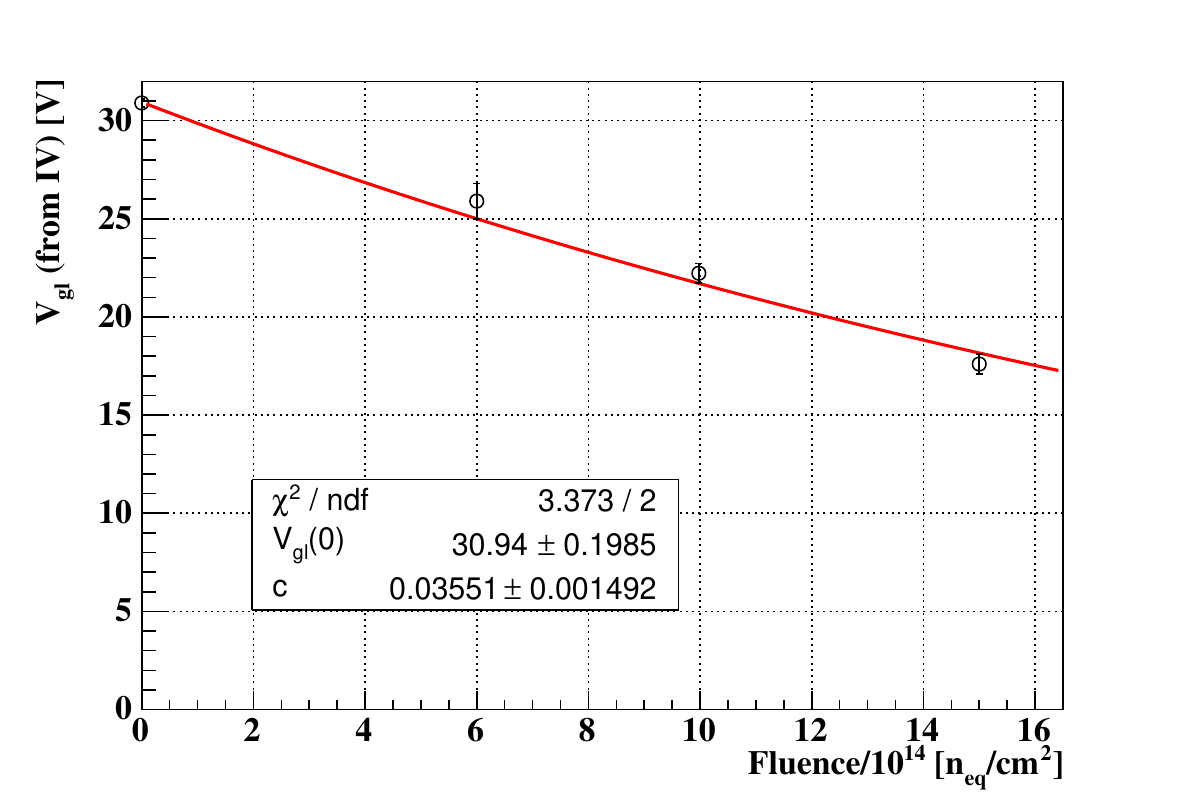}
         \caption{Carbonated samples}
     \end{subfigure}
     \hfill
     \begin{subfigure}[b]{0.43\textwidth}
         \centering
         \includegraphics[width=\textwidth]{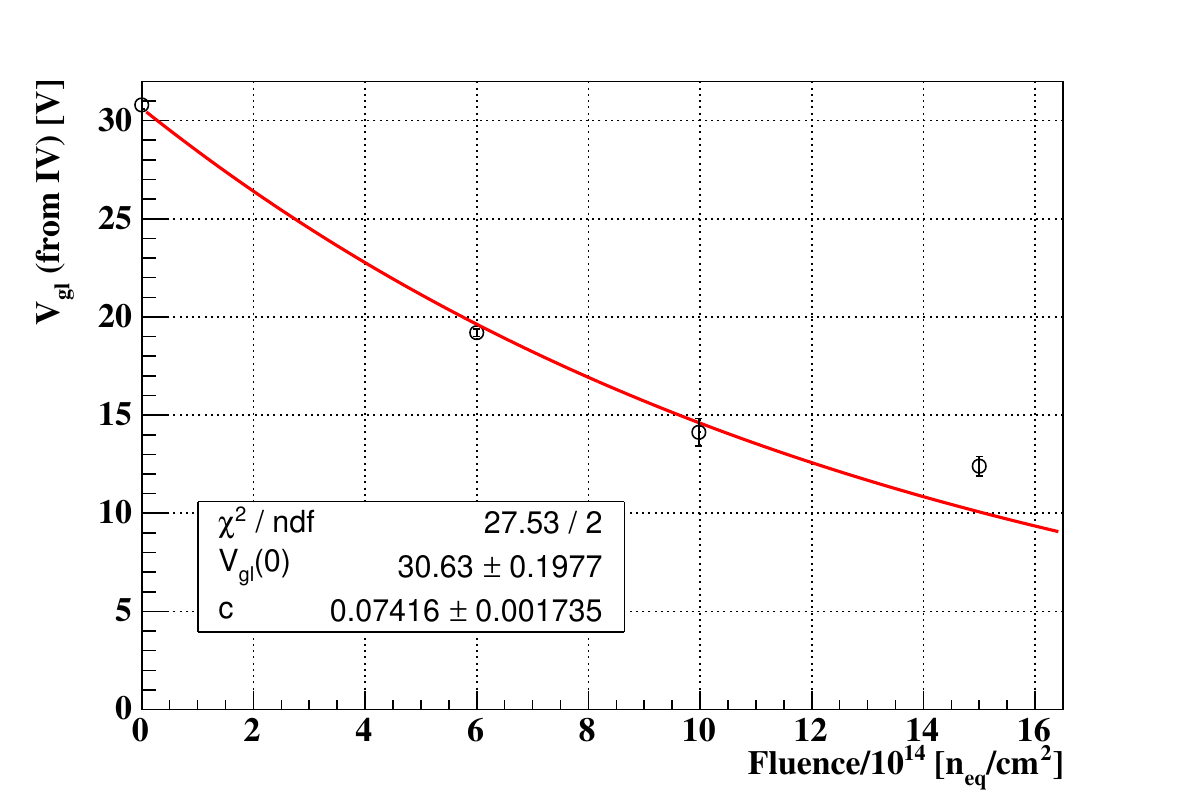}
         \caption{Standard samples}
     \end{subfigure}
     \begin{subfigure}[b]{0.43\textwidth}
         \centering
         \includegraphics[width=\textwidth]{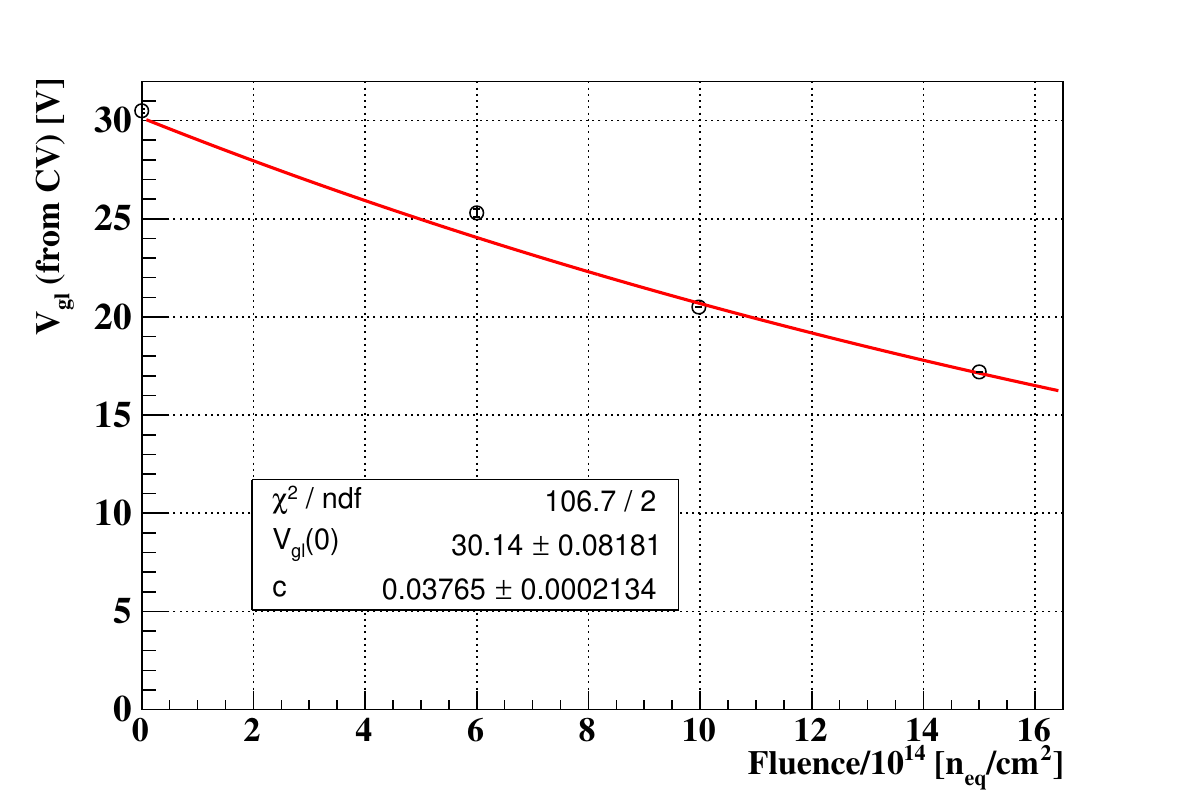}
         \caption{Carbonated samples}
     \end{subfigure}
     \hfill
     \begin{subfigure}[b]{0.43\textwidth}
         \centering
         \includegraphics[width=\textwidth]{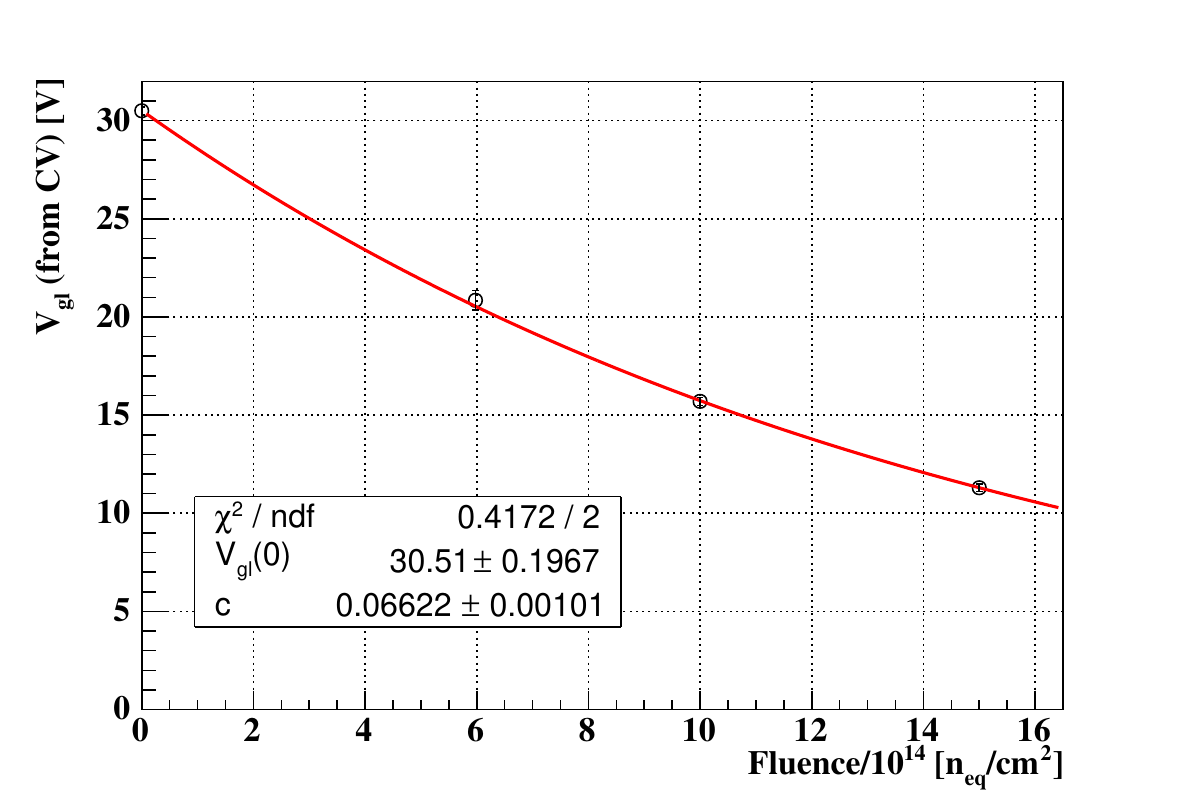}
         \caption{Standard samples}
     \end{subfigure}
        \caption{(a) and (b) contain the gain-layer depletion voltage from IV measurements and (c) and (d) from the CV measurements, both as a function of the fluence with their respective fit in red, from where the acceptor removal coefficient is calculated and being: $c[\SI{e-16}{\centi\meter\squared}]=3.6$ for carbonated samples (a) and $c[\SI{e-16}{\centi\meter\squared}]=7.4$ for the standard (b) from IV calculation, and $c[\SI{e-16}{\centi\meter\squared}]=3.8$ for carbonated samples (c) and $c[\SI{e-16}{\centi\meter\squared}]=6.6$ for the standard (d) from CV. The errors of the points are the standard error of the mean.}
        \label{fig:Acceptor}
\end{figure}

The resulting curve for $V_{GL}$ versus fluence, along with its fit, is presented in \autoref{fig:Acceptor} for both carbonated and standard devices. The resulting coefficients are $c[\SI{e-16}{\centi\meter\squared}] = 3.6$ and $c[\SI{e-16}{\centi\meter\squared}] = 7.4$, respectively, measured from the IV curves, and $c[\SI{e-16}{\centi\meter\squared}] = 3.8$ and $c[\SI{e-16}{\centi\meter\squared}] = 6.6$ for the carbonated and standard from the CV measurement. This result indicates that the addition of carbon in the co-implantation of the gain layer leads to a lower boron deactivation and, consequently, an improvement in the radiation tolerance of these LGADs. The determination of the acceptor removal coefficient, from IV or CV characterization is affected by different systematic effects difficult to account for precisely. To take them into account, we could average the acceptor removal coefficient obtained from IV and CV characterizations, the resulting coefficient for carbonated sensors is close to a half of the standard sensors.

\section{Radioactive Source Characterization}
\label{sec:RS}

\begin{figure}
     \centering
     \begin{subfigure}[b]{0.49\textwidth}
         \centering
         \includegraphics[width=\textwidth]{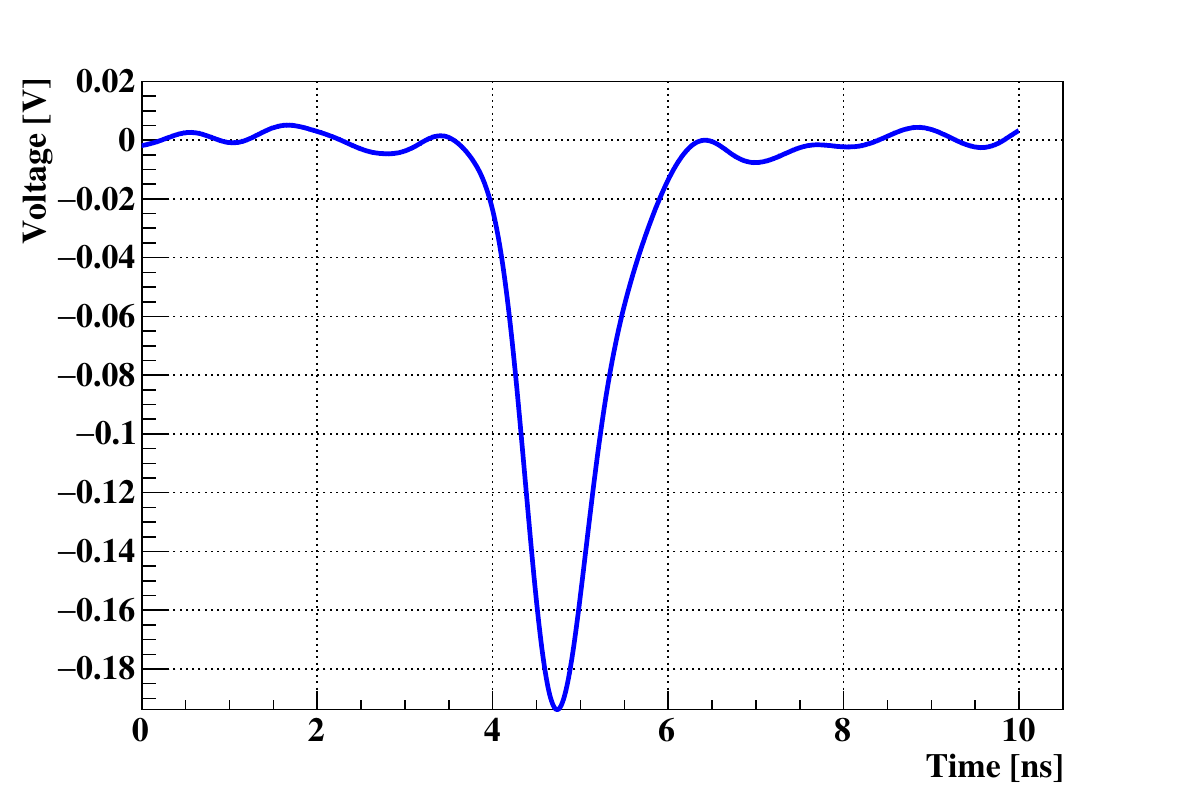}
         \caption{Typical waveform.}
     \end{subfigure}
     \hfill
     \begin{subfigure}[b]{0.49\textwidth}
         \centering
         \includegraphics[width=\textwidth]{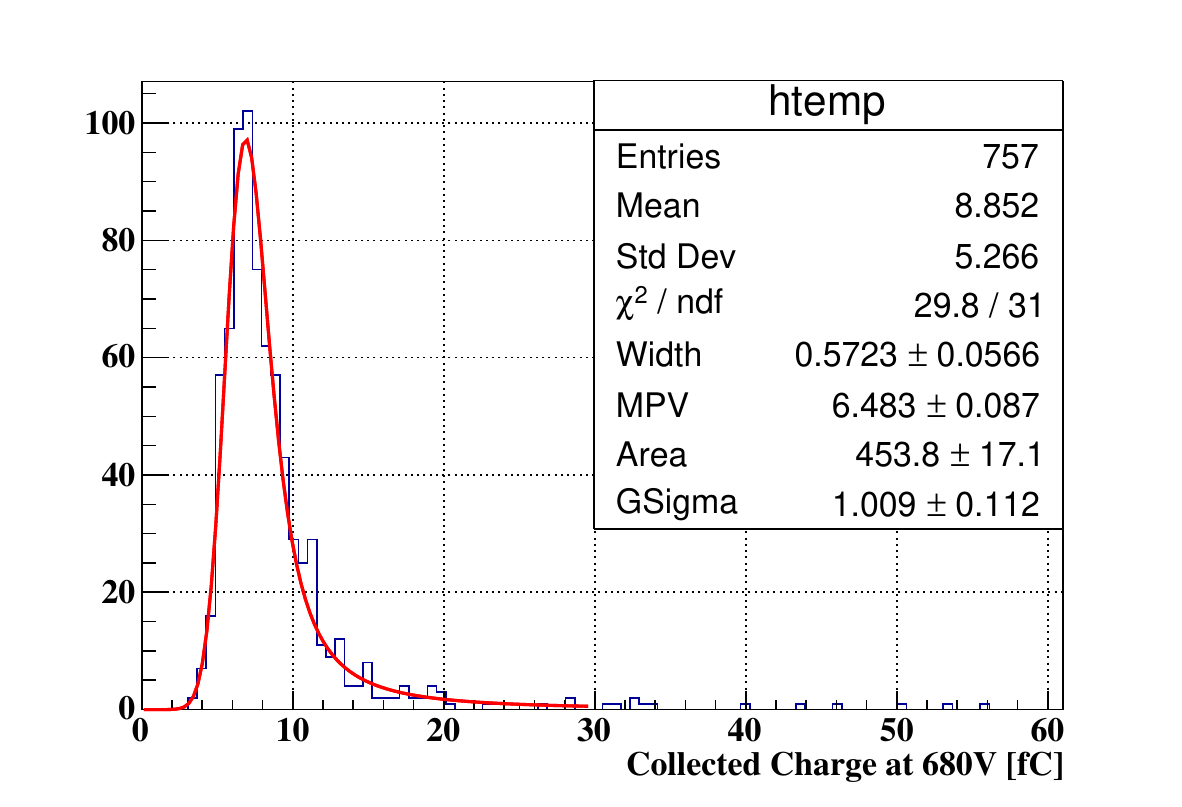}
         \caption{Collected charge.}
     \end{subfigure}
        \caption{(a) Typical waveform from an LGAD with the voltage response versus the time of the pulse from a non-irradiated LGAD biased at $\SI{260}{\volt}$ . (b) Distribution of the collected charge computed from the integration of the waveforms of a carbonated LGAD irradiated to $\fluenceUnits{15e14}$ and biased at $\SI{680}{\volt}$. From the convoluted Gauss-Landau fit the Most Probable Value (MVP) can be extracted.}
        \label{fig:chargeColl}
\end{figure}

The radioactive source setup at IFCA consists of a metallic box enclosing a stack of three sensors. Each sensor is mounted on a simple passive PCB that provides electrical connections. The box is placed inside a climate chamber that allows for temperature cycles. An encapsulated Sr$^{90}$ radioactive source with an activity of \SI{3.7}{\mega\becquerel} is positioned on top of the stack, ensuring no physical contact with the samples. The alignment of sensors in the stack is maintained by gluing the devices using a mechanical template. To measure the current induced, an external low-noise current amplifier (with a nominal gain of \SI{40}{\decibel})~\cite{CIVIDEC} is used for amplification. The readout is performed using an oscilloscope with a sampling rate of \SI{5}{\giga S\per\second}. The readout is triggered when a MIP is registered by the three sensors in the same single event, referred to here as a triple coincidence. The samples measured in the radioactive source setup are detailed in~\autoref{tab:rs-sensors}. The third detector in the stack was always a non-irradiated LGAD device, serving as a reference.

\subsection{Charge Collection}
\label{sec:Charge}

The collected charge per particle is calculated as the integral of the voltage pulse (\autoref{fig:chargeColl} (a)). The total distribution of charge for a single detector, shown in \autoref{fig:chargeColl} (b), is fitted by the convolution of a Landau with a gaussian. The most probable value of this distribution is used as an estimation of the total collected charge.

\begin{figure}
     \centering
     \begin{subfigure}[b]{0.49\textwidth}
         \centering
         \includegraphics[width=\textwidth]{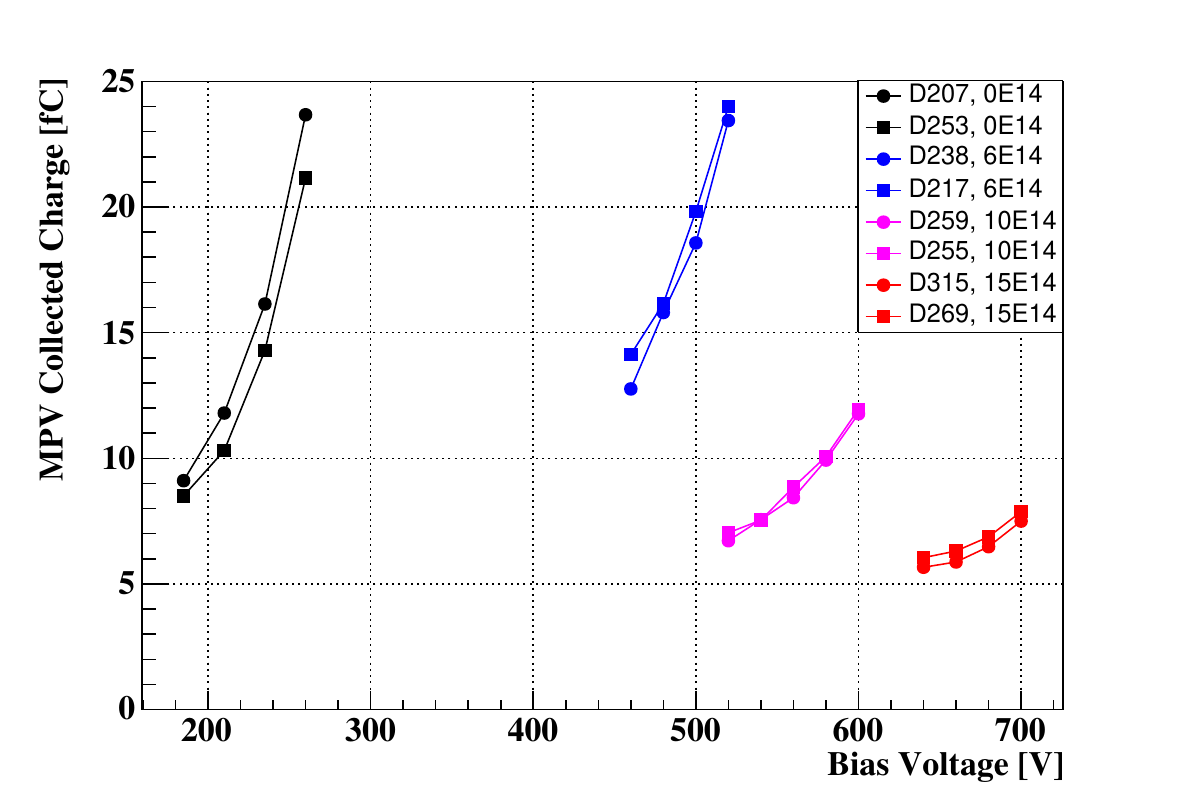}
         \caption{Carbonated LGADs}
     \end{subfigure}
     \hfill
     \begin{subfigure}[b]{0.49\textwidth}
         \centering
         \includegraphics[width=\textwidth]{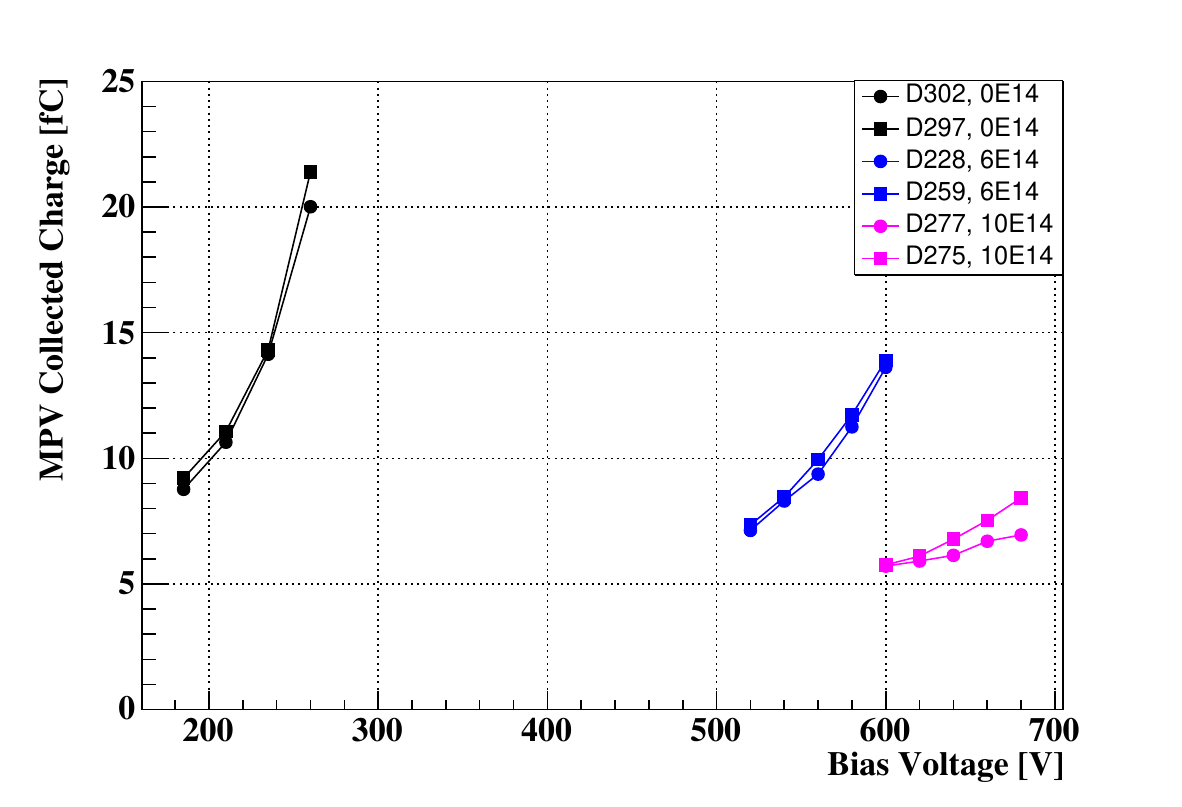}
         \caption{Standard LGADs}
     \end{subfigure}
        \caption{Collected Charge as a function of the reverse bias voltage for carbonated samples can be shown in (a) and samples with standard gain layer in (b). There is a difference in voltage required to achieve the same level of collected charge between the two types of LGADs at a same fluence value. All these measurements were performed at a temperature of \SI{-25}{\celsius}.}
        \label{fig:Charge}
\end{figure}

\autoref{fig:Charge} shows the charge collected as a function of the bias voltage for different fluences. There is a clear separation between the samples in terms of the reverse bias regions. To achieve the same collected charge, the higher irradiated samples require higher bias. There is a good repeatibility of the collected charge for samples irradiated to the same fluence. The carbonated sensors can be operated at a bias lower than the standard ones. For example, comparing carbonated and non-carbonated sensors irradiated to the same fluence of \fluenceUnits{6e14} to collect $\SI{14}{\femto\coulomb}$, $\SI{460}{\volt}$ is required for a carbonated sample, whereas $\SI{600}{\volt}$ is required for a non-carbonated sample, indicating a higher radiation-induced degradation for the non-carbonated sample, which requires around \SI{140}{\volt} more bias voltage to collect the same amount of charge than the carbonated.

\subsection{Time Resolution}
\label{sec:time}

The time resolution of a sensor can be calculated as the standard deviation of the distribution of the differences in arrival time (ToA) of the sensor with respect to a well-known reference. For these cases where no time reference is available, three detectors can be measured simultaneously~\cite{McKarris} and the individual time resolutions calculated from the three relative differences (1-2, 1-3, 2-3). 

The time of arrival (ToA) is computed as the time when a pulse crosses a threshold. Since pulses of different amplitudes arriving at the same time will cross a threshold at different times (time walk effect), the pulses are corrected using a Constant Fraction Discrimination (CFD) algorithm.

The fitted widths: $\sigma_{1,2}$, $\sigma_{1,3}$, and $\sigma_{2,3}$ of the difference distributions are used to determine the time resolution of the three sensors ($\sigma_{1}$, $\sigma_{2}$, $\sigma_{3}$) by solving the system of equations:

\begin{equation}
\label{eq:sigmas}
\begin{split}
\sigma_1= \left( \frac{1}{2} (\sigma_{2,1}^{2}+\sigma_{1,3}^{2}-\sigma_{3,2}^{2})\right)^{\frac{1}{2}} \,, \\
\sigma_2= \left(\frac{1}{2} (\sigma_{2,1}^{2}-\sigma_{1,3}^{2}+\sigma_{3,2}^{2})\right)^{\frac{1}{2}} \,, \\
\sigma_3= \left(\frac{1}{2} (-\sigma_{2,1}^{2}+\sigma_{1,3}^{2}+\sigma_{3,2}^{2})\right)^{\frac{1}{2}} \,, 
\end{split}
\end{equation}

with errors ($\delta_{1}$, $\delta_{2}$ and $\delta_{3}$) given by:

\begin{equation}
\label{eq:deltas}
\begin{split}
\delta_1= \frac{ \left( (\sigma_{2,1} \delta_{2,1})^2 + (\sigma_{1,3} \delta_{1,3})^2 + (\sigma_{3,2} \delta_{3,2})^2\right)^{\frac{1}{2}} }{2\sigma_1}\,,\\
\delta_2= \frac{ \left( (\sigma_{2,1} \delta_{2,1})^2 + (\sigma_{1,3} \delta_{1,3})^2 + (\sigma_{3,2} \delta_{3,2})^2\right)^{\frac{1}{2}} }{2\sigma_2}\,,\\
\delta_3= \frac{ \left( (\sigma_{2,1} \delta_{2,1})^2 + (\sigma_{1,3} \delta_{1,3})^2 + (\sigma_{3,2} \delta_{3,2})^2\right)^{\frac{1}{2}} }{2\sigma_3}\,,
\end{split}
\end{equation}

where $\delta_{i,j}$ is the error in the value $\sigma_{i,j}$.

\begin{figure}
     \centering
     \begin{subfigure}[b]{0.49\textwidth}
         \centering
         \includegraphics[width=\textwidth]{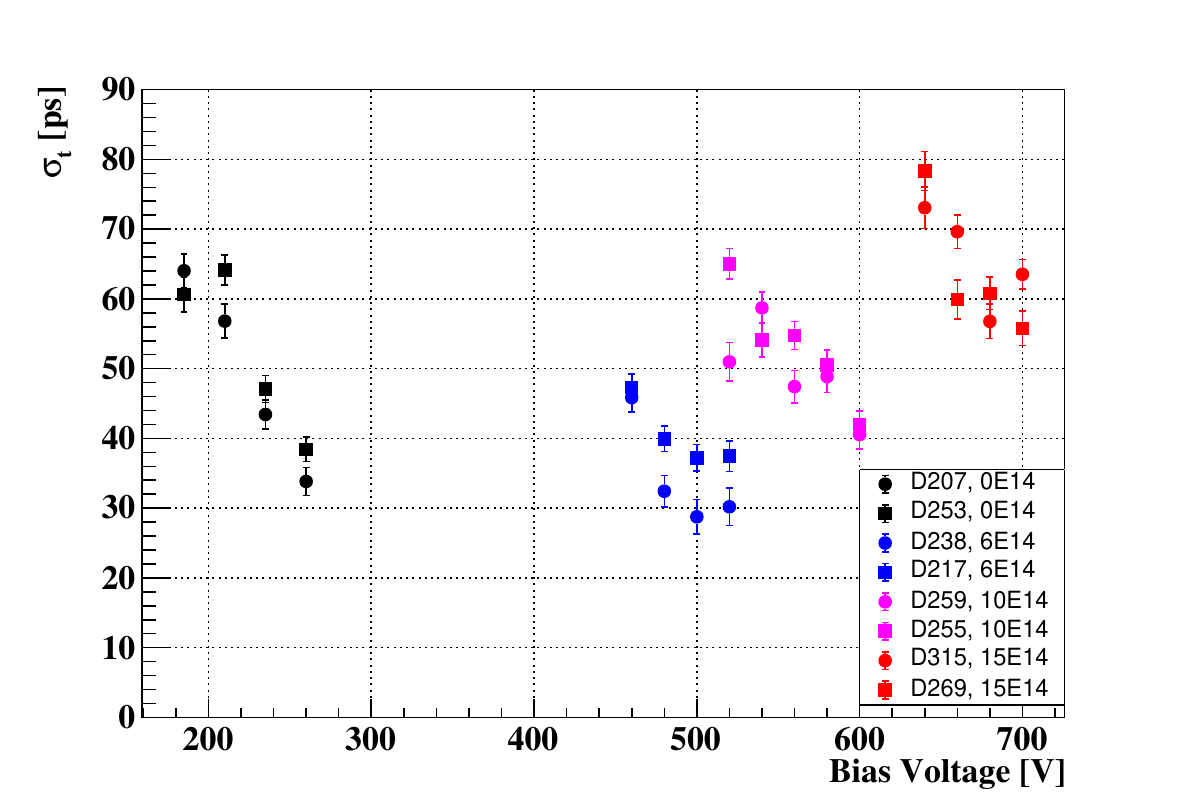}
         \caption{Carbonated LGADs}
     \end{subfigure}
     \hfill
     \begin{subfigure}[b]{0.49\textwidth}
         \centering
         \includegraphics[width=\textwidth]{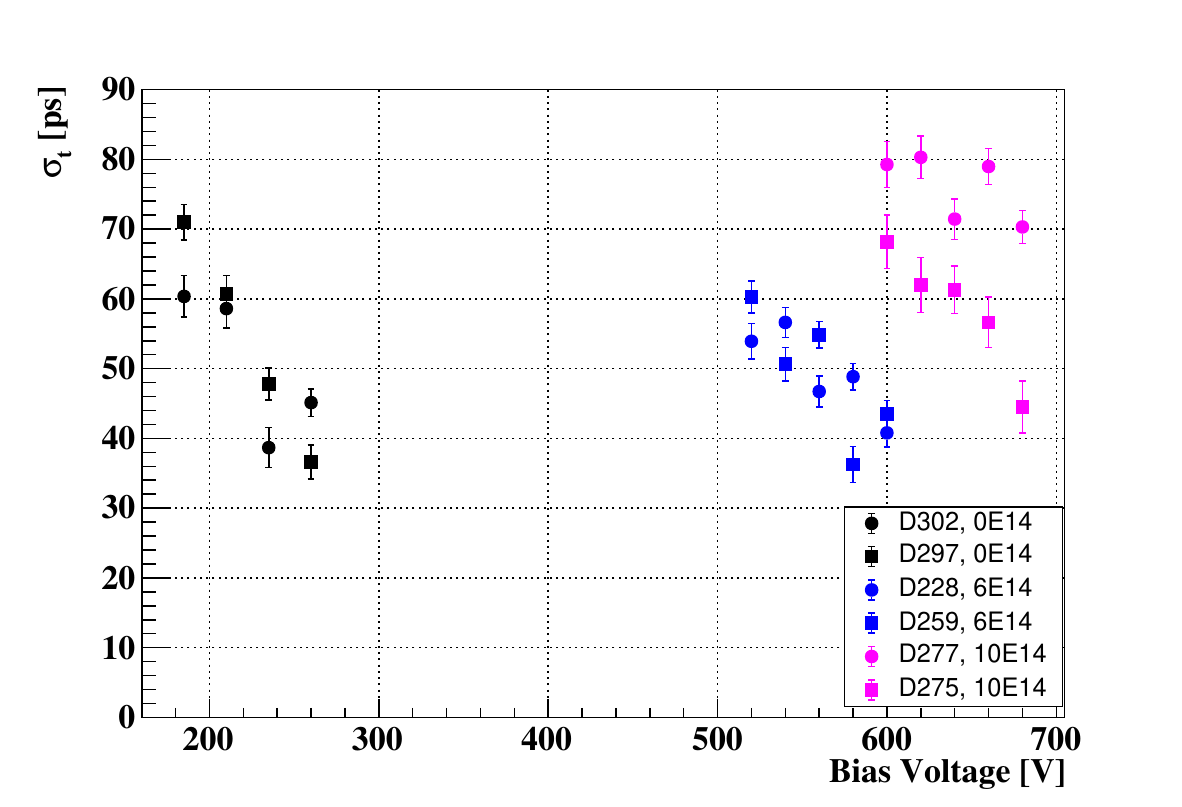}
         \caption{Standard LGADs}
     \end{subfigure}
        \caption{Time resolution of both type of sensors: Carbonated (a) and Standard (b), calculated using \autoref{eq:sigmas} and with errors calculated with \autoref{eq:deltas}. The carbonated sensors show a better behavior after irradiation compared to the standard sensors. All these measurements were performed at a temperature of $\SI{-25}{\celsius}$.}
        \label{fig:Time}
\end{figure}

This method was applied to all samples of this study, maintaining the non-irradiated sensor mentioned in~\autoref{sec:RS} as reference. The resulting time resolution $\sigma_t$ of carbonated and standard sensors is presented in \autoref{fig:Time} (a) and (b) respectively. As the fluence increases, the voltage needed to achieve the same time resolution also increases. Time resolution improves as bias voltage grows. Finally, the bias voltage needed to reach values below \SI{50}{\pico\second} of time resolution is about \SI{100}{\volt} smaller in carbonated detectors after irradiation.

\section{Noise Study}
\label{sec:noise}

To ensure the functionality of the sensors, a noise study was carried out on the carbonated samples. Although this noise affects both carbonated and non-carbonated samples, this study was not carried out on the standard sensors. Key parameters considered include the baseline noise level and the presence and frequency of micro-discharges (thermally generated spurious pulses).

\begin{figure}
     \centering
     \begin{subfigure}[b]{0.42\textwidth}
         \centering
         \includegraphics[width=\textwidth]{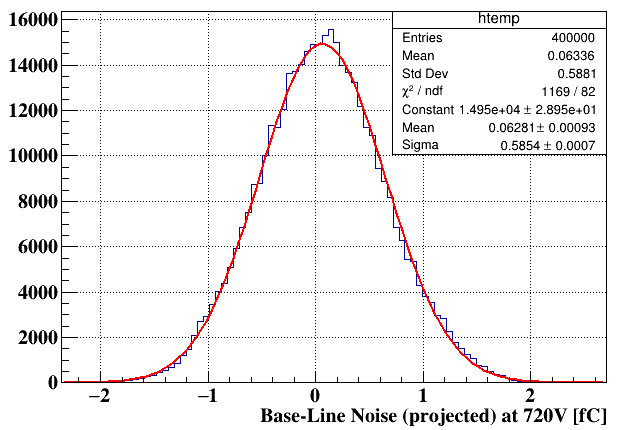}
         \caption{Base-Line distribution example}
     \end{subfigure}
     \hfill
     \begin{subfigure}[b]{0.49\textwidth}
         \centering
         \includegraphics[width=\textwidth]{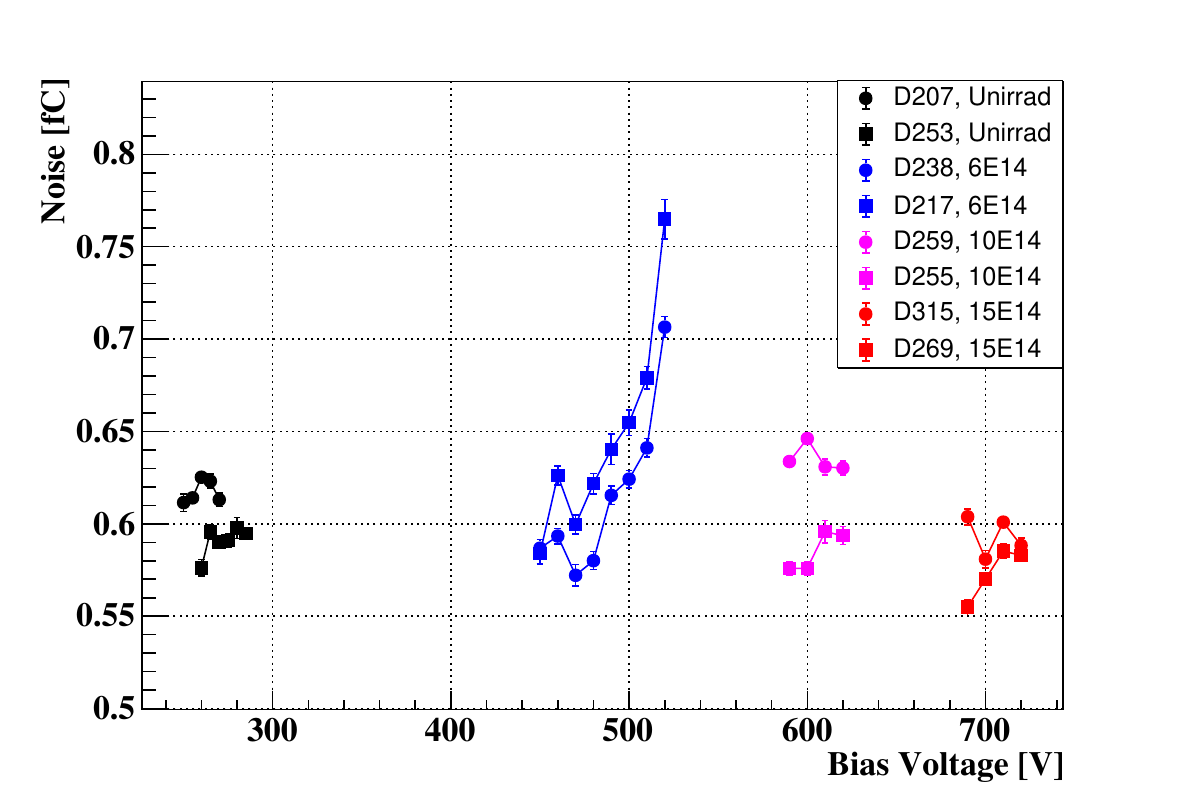}
         \caption{Base-Line Noise versus bias voltage}
     \end{subfigure}
     \caption{Example of the distribution of the signals response from a carbonated LGAD irradiated to $\fluenceUnits{1.5e15}$ without Radioactive Source in order to extract the Base-Line noise and its gausian fit can be seen in (a). (b) is the summary of the extracted Base-Line Noise of the carbonated samples at the different fluences versus the bias voltage.}
     \label{fig:BaseLineNoise}
\end{figure}

The noise of the carbonated samples was investigated using a random trigger. Two hundred events (waveforms) per bias voltage were collected in a histogram (depicted in \autoref{fig:BaseLineNoise} (a)), and the Gaussian width of the resulting distribution was taken as the noise value. The increase in noise with voltage was examined for various fluence values and summarized in \autoref{fig:BaseLineNoise} (b). The increase in noise is more pronounced for the \fluenceUnits{0.6e15} samples, but does not prevent the sensors from reaching operating conditions, that is, the reverse current remains low enough. The fact that the \fluenceUnits{0.6e15} samples present a higher noise pulse rate may result to the fact that at this fluence point the gain suppression is relatively smaller than in the sensors exposed to higher fluences; therefore having a larger excess noise. The measurements were performed at a temperature of $\SI{-25}{\celsius}$ but the study of noise at different temperatures will be carried out in  future studies.

\begin{figure}
     \centering
         \centering
         \includegraphics[width=0.5\textwidth]{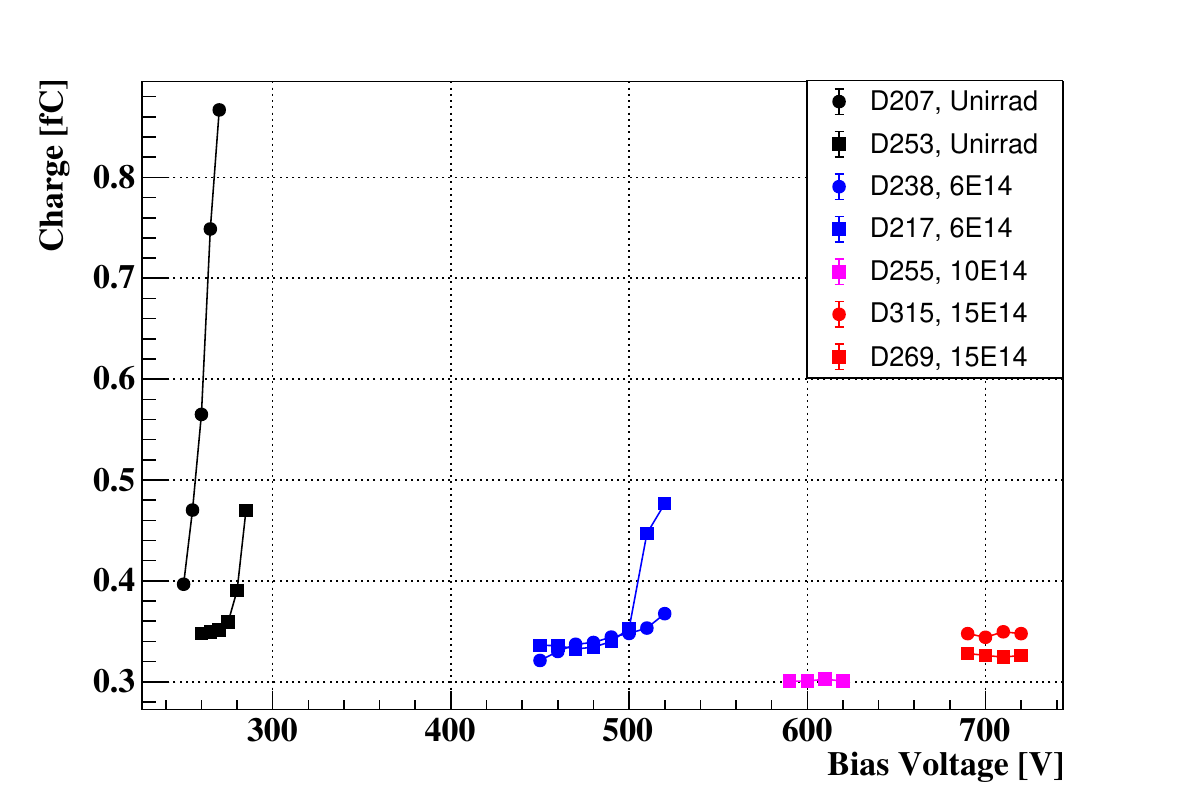}
        \caption{Collected charge of the spurious pulses from the carbonated samples irradiated and non-irradiated. Calculated from the amplitude-charge correlation.  Measurements taken at a temperature of \SI{-25}{\celsius}.}
        \label{fig:SpuriousAmplitude}
\end{figure}

The amplitude of dark counts or spurious pulses that appear when a high electric field induces micro-avalanches triggered by thermal generated primary electrons, was also measured at voltages close to breakdown for each fluence with a trigger of threshold level of $\SI{-15}{\milli\volt}$, this threshold corresponds to approximately 6 MIPs for a PIN diode (without gain) of \SI{50}{\micro\meter} of active thickness.
Again, no radioactive source was employed for this study. The spurious pulses appeared in all the carbonated samples near the breakdown voltage. \autoref{fig:SpuriousAmplitude} shows the equivalent collected charge for these spurious pulses, calculated from the amplitude-charge correlation obtained from the measurements taken in~\autoref{sec:RS}. For the non-irradiated and lowest fluence samples, a scaling trend with the bias voltage can be seen, while for the two higher fluence samples no increase was observed near breakdown.

\begin{figure}
     \centering
     \begin{subfigure}[b]{0.49\textwidth}
         \centering
         \includegraphics[width=\textwidth]{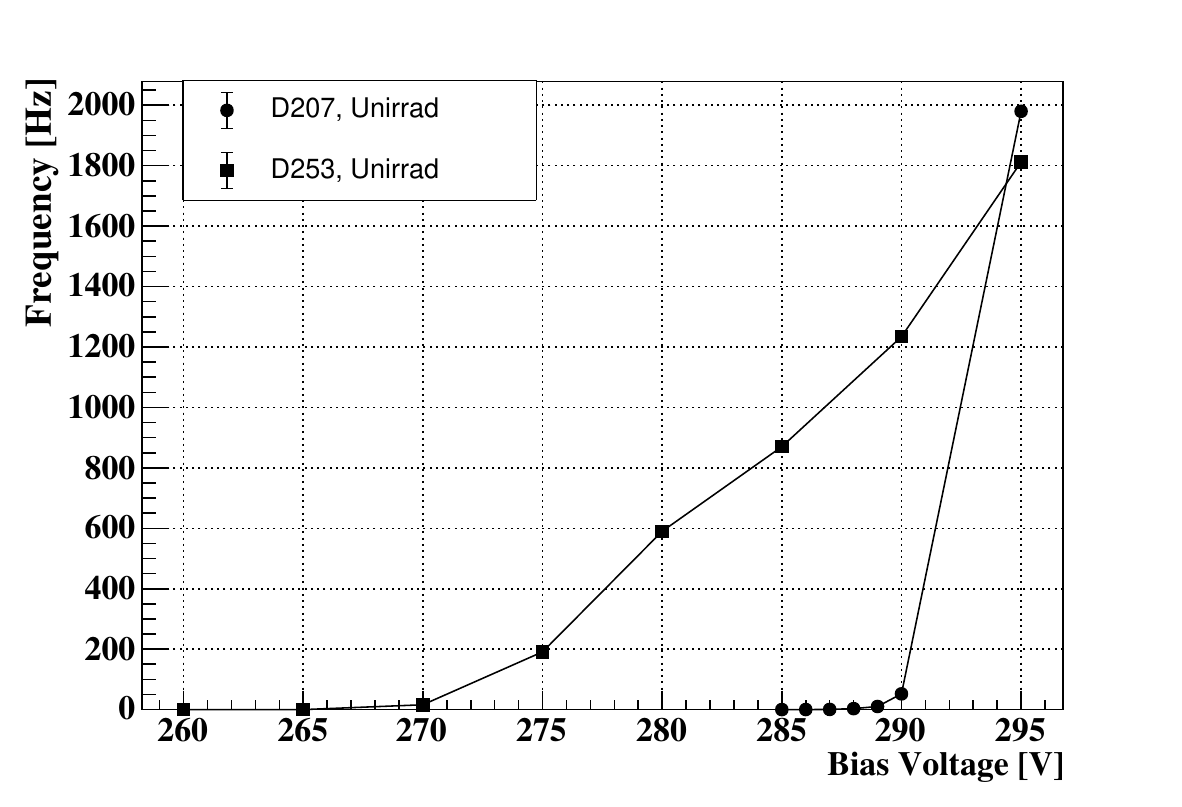}
         \caption{Non-irradiated}
     \end{subfigure}
     \hfill
     \begin{subfigure}[b]{0.49\textwidth}
         \centering
         \includegraphics[width=\textwidth]{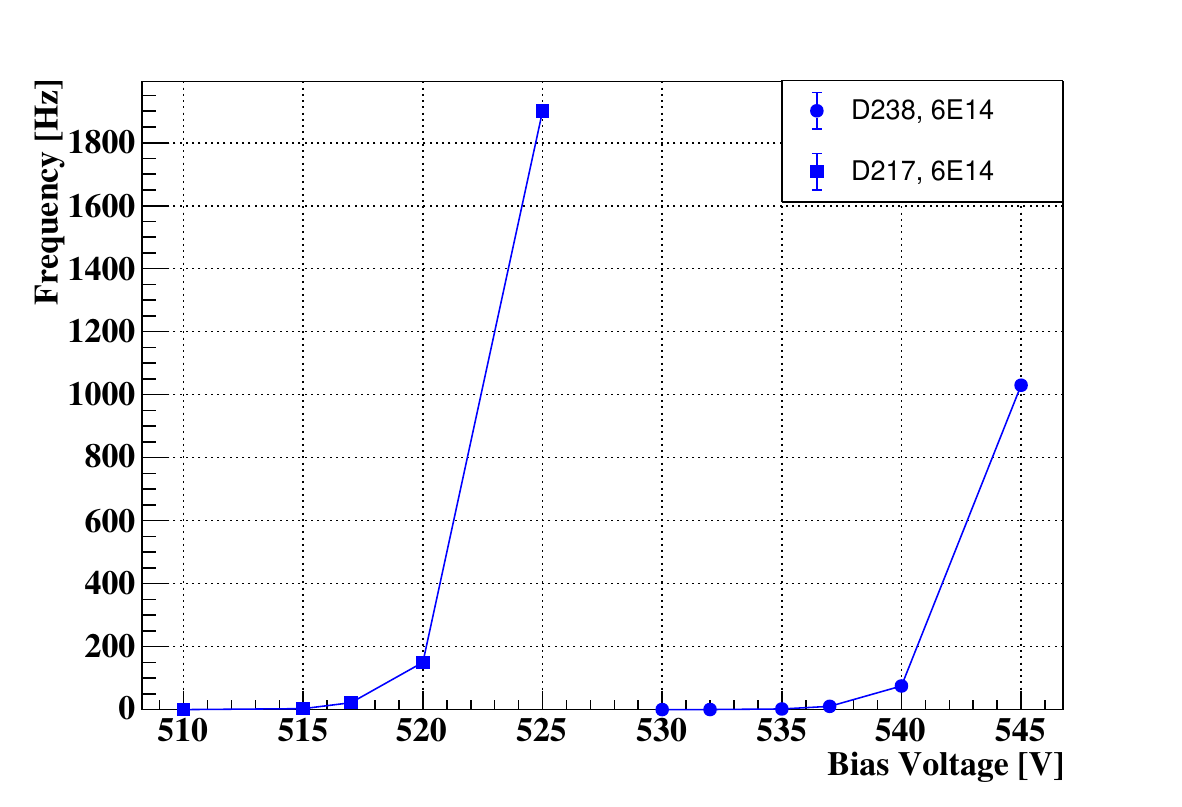}
         \caption{$\fluenceUnits{0.6e15}$}
     \end{subfigure}
     \begin{subfigure}[b]{0.49\textwidth}
         \centering
         \includegraphics[width=\textwidth]{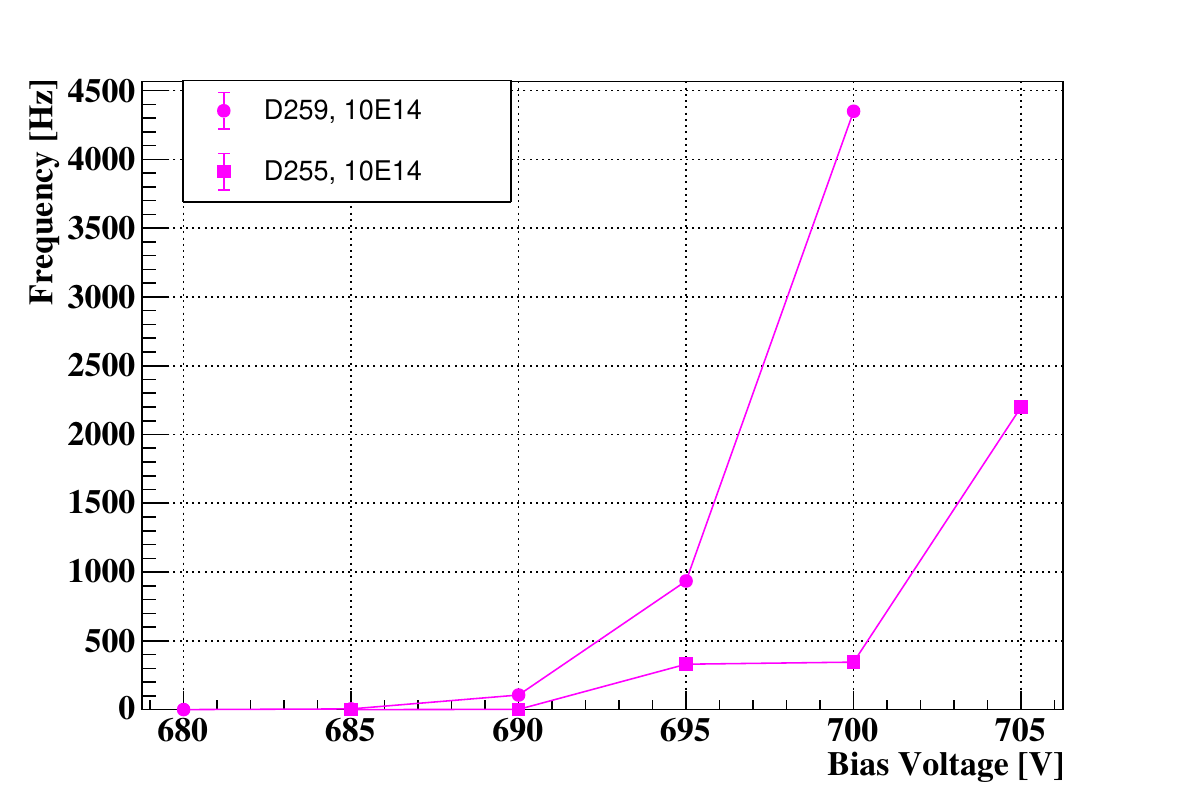}
         \caption{$\fluenceUnits{1.0e15}$}
     \end{subfigure}
     \hfill
     \begin{subfigure}[b]{0.49\textwidth}
         \centering
         \includegraphics[width=\textwidth]{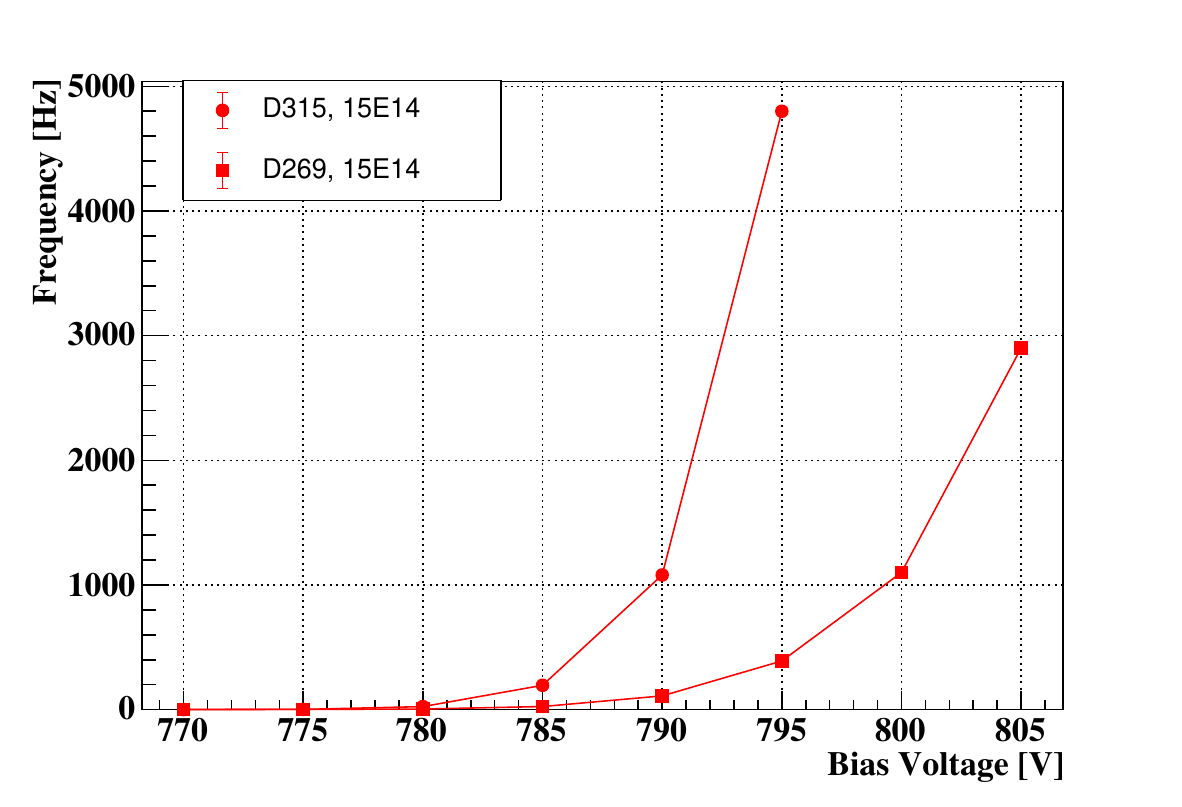}
         \caption{$\fluenceUnits{1.5e15}$}
     \end{subfigure}
        \caption{Spurious pulse rate versus the bias voltage of the carbonated samples before irradiation (a), and at $\fluenceUnits{0.6e15}$ (b), $\fluenceUnits{1.0e15}$ (c) and $\fluenceUnits{1.5e15}$ (d) irradiation fluences. Measurements taken in the Radioactive Source setup with NIM electronics with a threshold of $\SI{-25}{\milli\volt}$.}
        \label{fig:SpuriousPulseRates}
\end{figure}

To avoid the pulse rate being limited by the bandwidth of the digital scope, we decided to use NIM~\cite{NIM} electronic modules, specifically a Discriminator, a Timer and a Counter to obtain the pulse rate of the Dark Counts. The minimum threshold of the discriminator was $\SI{-25}{\milli\volt}$ that is higher than the $\SI{-15}{\milli\volt}$ used with the oscilloscope, resulting in an underestimation of the spurious pulse rate. The resulting rates for the different carbonated samples are shown in \autoref{fig:SpuriousPulseRates}. 

The plot starts at the operating voltage for each detector, that is the bias voltage needed to obtain $\SI{8}{\femto\coulomb}$ by a MIP (the so-called V(\SI{8}{\femto\coulomb}) requirement for the ETL~\cite{Market-Survey}): $\SI{240}{\volt}, \SI{460}{\volt}, \SI{580}{\volt}$ and $\SI{690}{\volt}$ respectively for the non-irradiated, $\fluenceUnits{0.6e15}$, $\fluenceUnits{1.0e15}$ and $\fluenceUnits{1.5e15}$ irradiated samples. The early presence of spurious pulses was attributed to the short JTE and the distance between the end of the $p^{+}$ gain layer and the p-stop at the periphery of the pad (see \autoref{fig:scheme}). For this production this distance was \SI{23.5}{\micro\meter}, which was decided in order to minimize, as much as possible, the inter-pad distance~\cite{periphery}, which is one of the main challenges for this technology in multi-pad matrix type LGADs.

\section{Conclusions}
\label{sec:Conclusions}

In this study, the first manufacturing run at IMB-CNM of Low Gain Avalanche Detectors with a carbon-enriched multiplication layer was investigated for its radiation tolerance compared to conventional LGADs. The sensors were subjected to neutron irradiation at the TRIGA reactor in Ljubljana, reaching a fluence of \fluenceUnits{1.5e15}. The results, reported in terms of degradation in timing performance and charge collection with increasing fluence, demonstrated the potential benefits of carbon enrichment in mitigating radiation damage effects, particularly the acceptor removal mechanism. The acceptor removal constant of the carbonated samples with respect to the standard samples was reduced to less (more) of a half from the results of the IV (CV) curves.

Time resolution and the collected charge was studied on the Radioactive Source (RS) setup for samples non-irradiated and irradiated up to fluences of  $\fluenceUnits{1.5e15}$. As expected, degradation of the time resolution and the collected charge due to the irradiation was evidenced. The time resolution of the carbonated samples, at a fluence of $\fluenceUnits{10e14}$ at the maximum bias voltage of $\SI{600}{\volt}$ achieved before the breakdown regime, is of $\SI{40}{\pico\second}$ while at same fluence and bias voltage for the Standard LGADs is about $\SI{75}{\pico\second}$, and at a maximum bias voltage of $\SI{680}{\volt}$ is around $\SI{57}{\pico\second}$. In terms of radiation tolerance, the carbonated samples meet the CMS ETL requirements for the low fluence region  (below \fluenceUnits{10e14}), with a time resolution of less than \SI{50}{\pico\second} with a collected charge of \SI{8}{\femto\coulomb}.

Additionally, a  noise analysis was conducted on the samples. The investigation focused on key parameters, including baseline noise level and the occurrence and frequency of micro-discharges, which may manifest as spurious pulses in silicon detectors due to thermal generation. The noise of carbonated samples was analyzed using a random trigger, measuring signal width without a radioactive source. The resulting noise values were examined across various fluences, as depicted in \autoref{fig:SpuriousAmplitude}. Despite a more pronounced increase in noise for samples irradiated to $\fluenceUnits{0.6e15}$, the elevated noise levels did not impede the device's operation. Additionally, spurious, thermally generated pulses were measured beyond the operational voltages, showing a scaling trend with bias voltage for non-irradiated and lower fluence samples, while higher fluence samples exhibited no increase near breakdown.

\section*{Acknowledgments}

This work was developed in the framework of the CERN RD50/DRD3 collaboration and has been funded by the Spanish Ministry of Science and Innovation (MCIN/AEI/10.13039/501100011033/) and by the European Union’s ERDF program “A way of making Europe”. Grant references: PID2020-113705RB-C31, PID2020-113705RB-C32 and PID2021-124660OB-C22. Also, it was supported by the following European funding programs: the European Union’s Horizon 2020 Research and Innovation (under Grant Agreement No. 101004761, AIDAInnova) and NextGenerationEU (PRTR-C17.I1).
This work was also been developed in the framework of the "Ayudas Maria Zambrano para la atraccion de talento internacional", co-funded by the Ministry Of University of Spain and the European Union NextGenerationEU, reference code: C21.I4.P1;  under the framework of "Ayudas para contratos predoctorales para la formación de doctores 2019", co-funded by the European Social Fund program "El FSE invierte en tu futuro" with grant reference: PRE2019-087514; and the  \emph{Plan Complementario en el Área de Astrofísica y Física de Altas Energías}, financed by Next Generation EU funds, including the Recovery and Resilience Mechanism (MRR), the Recovery, Transformation and Resilience Plan (PRTR) and the Autonomous Community of Cantabria.

\end{document}